\documentclass[12pt]{article}
\pdfoutput=1 
\usepackage[utf8]{inputenc}
\usepackage{indentfirst}
\usepackage{amsmath,amsthm,amssymb}
\usepackage{graphicx}
\usepackage{setspace}
\usepackage{hyperref}
\usepackage[all]{hypcap}
\usepackage{lipsum}

\title{The dilute Temperley--Lieb O($n=1$) loop model on a semi infinite strip: the ground 
state}

\author{A. Garbali$^{1,2,3}$ and B. Nienhuis$^{2}$
\vspace{0.5cm}
\\
$^{1}$LPTHE, CNRS UMR 7589, \\ 
Universit\'e Pierre et Marie Curie (Paris 6),\\
4 place Jussieu, 75252 Paris cedex 05, France.
\and
$^{2}$Instituut voor Theoretische Fysica,\\
Universiteit van Amsterdam,
Science Park 904, \\
1098 XH Amsterdam, The Netherlands.
\and
$^{3}$ARC Centre of Excellence for Mathematical and Statistical Frontiers (ACEMS), \\School of Mathematics and Statistics, University of Melbourne, \\Parkville, Victoria 3010, Australia.}
\date{}
\begin{document}
\maketitle

\begin{abstract}
We consider the integrable dilute Temperley--Lieb (dTL) O($n=1$) loop model on 
a semi-infinite strip of finite width $L$.
In the analogy with the Temperley--Lieb (TL) O($n=1$) loop model the ground state eigenvector 
of the transfer matrix is studied by means 
of a set of $q$-difference equations, sometimes called the $q$KZ equations. We compute some 
ground state components of the transfer matrix of the dTL model, and show that all ground state 
components can be recovered for arbitrary $L$ using the $q$KZ equation and certain recurrence 
relation.  The computations are done for generic open boundary conditions.
\end{abstract}

\section{Introduction}
In the last decade the integrable loop models with the loop weight $n=1$ became a subject of 
great 
interest due to their relation to combinatorics, algebraic geometry, percolation, etc. 
Probably, the most striking result is the Razumov--Stroganov (RS) conjecture, relating the 
components of the ground state of the Temperley-Lieb (TL) loop model to  
alternating sign matrices \cite{BGN,RS}.  The weaker version of this conjecture \cite{BGN} was 
proved in \cite{PDFPZJ}, while the stronger RS conjecture \cite{RS} in \cite{CS}. 
The ground state of the TL loop model is known to be related to certain 
orbital varieties.
Namely, the entries of the ground state of the TL loop model coincide with the multidegrees 
of certain matrix varieties \cite{PDFPZJ,AKPZJ}.
The TL loop model at $n=1$ is equivalent to critical bond percolation. 
Studying the ground state of the TL model with 
different boundary conditions allows 
to compute some correlation functions for critical bond percolation 
\cite{MitraNienhuis,MitraNienhuis2}. 
Since the O($1$) TL loop model appears to be very 
fruitful it motivates us to study also the dilute Temperley-Lieb (dTL) loop model \cite{Nienhuis0,Nienhuis1}.

The dilute Temperley-Lieb (dTL) loop model, with loop weight $n=1$,  is a formulation of  
critical site percolation.  Hence, its ground state contains the information of 
certain correlation functions for critical site percolation. 
At $n=0$ the ground state of the dTL model may provide a method to approach the 
closed self avoiding walk in 2D. In both cases ($n=1$ and $n=0$) the scaling limits are 
described by a Schramm--Loewner evolution (SLE). It is, however, a conjecture for the self 
avoiding walk.  The dTL loop model at finite size may provide insights to the relations to SLE.

We organize the paper as follows. In Section \ref{sec2} we present the definitions and 
the basic ingredients. In Section \ref{sec3} we set the loop weight to $1$ and study the $q$KZ 
equations, after we discuss the recurrence in size and finally compute the ground state. 
Discussions follow in Section \ref{sec4}.  The explicit results for $\Psi_L$ for $L=1$ and $2$ 
are given in Appendix \ref{appA}.  The discussion of the recurrence relation is presented in Appendix \ref{appB} . 

\section{The dilute \texorpdfstring{O$(n)$}{Lg} loop model}\label{sec2}
The loop model we discuss here is defined on the square lattice with  finite 
width and infinite length.  This domain can be seen as a half infinite 
strip that has two vertical half infinite boundaries and one finite horizontal 
boundary of length $L$.  If we identify the vertical boundaries, and otherwise leave them free, this will lead to periodic 
boundary conditions.  Here we do not identify the vertical boundaries, but allow there any configuration, leading to open boundary conditions. 
The configurations of the dTL model are collections of non-intersecting paths over the edges of the lattice.  The paths are either closed, hence {\em loops}, or terminate on one of the boundaries.  Classes of configurations 
 can be labelled by the 
connectivities 
of the links on the horizontal boundary. Each edge of the horizontal boundary 
can be in two states: occupied or unoccupied. An occupied edge is connected to another 
edge on the horizontal boundary.  This way each configuration of the model corresponds 
to a link pattern on the horizontal boundary. All link patterns form a vector space lp$_L$, 
thus the states of the model can be expressed in this vector space. 
An important idea that has led to many advances for the TL model is to consider 
more general lattices, i.e. lattices with distortions or inhomogeneities $z_1,..,z_L$. 
When the loop weight is equal to $1$, the ground state 
configuration $\Psi_L$ becomes 
a vector with entries polynomial in $z_i$  satisfying certain $q$-difference equations, 
sometimes referred to as the quantum Knizhnik--Zamolodchikov equations ($q$KZ). 
A crucial ingredient for the computations is 
the conjectural expression for a certain entry of the ground state. Using this 
conjecture all other elements follow from the $q$KZ equations.  This was the basis 
of the algorithm 
for the computation of the ground state components and the sum rules for the TL O$(1)$ 
loop models with various boundary conditions \cite{PDFPZJ,PDF1,PZJ,dGPS,Ca} and for the 
dTL O($1$) model with periodic boundary conditions \cite{PDF}. We extend these results 
to the dTL O($1$) model with open boundary conditions.

\subsection{Link pattern basis and the \texorpdfstring{$R$}{Lg}-matrix}
\begin{figure}[htb]
\centering
\includegraphics[width=0.4\textwidth]{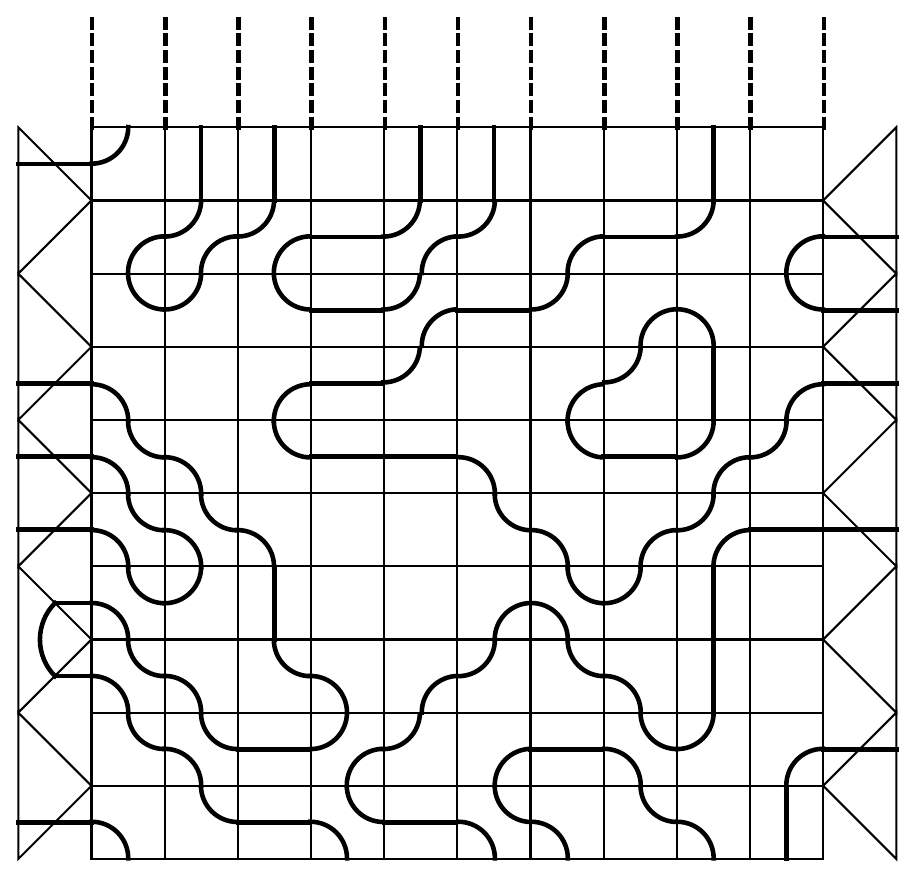}
\caption{A typical configuration of the dilute O$(n)$ loop model on a fragment of a 
semi-infinite strip.}
\label{figlp}
\end{figure}

\begin{figure}[htb]
\centering
\includegraphics[width=0.4\textwidth]{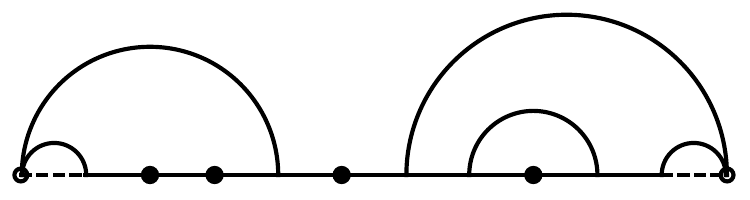}
\caption{The link pattern that corresponds to the configuration from Fig.  \ref{figlp}.}
\label{figlinkp}
\end{figure}

Consider the dilute O$(n)$ loop model on the square lattice on a semi-infinite strip. 
An example of a configuration of the model is given in Fig.  \ref{figlp}.
The paths can end at the horizontal boundary and two vertical boundaries. We classify all configurations according to their connectivity on the horizontal 
boundary, such a connectivity is called a link pattern. More precisely, let us 
label the edges of the lower boundary with numbers $1,2,..,L$ from left to right. If a 
path ends at an edge $i$, then this edge is said to be occupied, 
otherwise
the edge is unoccupied. Moreover, we will distinguish 
the occupied edges connected with other edges on their left and 
on their right.  
It is convenient to label the connectivity by the variables $n_i$ for
each edge $i$ which take the values $0$, $-1$ or $1$.
In a 
given configuration we assign the label $n_i=0$ to an unoccupied edge $i$.
We associate $n_i=1$ ($n_i=-1$) to an edge $i$ occupied and connected to an edge or a boundary on its right (left).
With this 
notation the connectivity of a configuration is a sequence $\{n_1,n_2,..,n_L\}$. For example, 
the configuration depicted in Fig. \ref{figlp} has the connectivity 
$\{-1,0,0, -1,0, 1, 1, 0, -1,1\}$. We can represent this pictorially as 
in Fig.  \ref{figlinkp}, where the vertical boundaries are represented by the leftmost 
and the rightmost points separated by a dashed line from the actual extremal edges  of the horizontal boundary.
The set of all link patterns, denoted by lp$_L$, is the basis in which we express 
the eigenvectors of the transfer matrix of the model.

\begin{figure}[htb]
\centering
\includegraphics[width=0.7\textwidth]{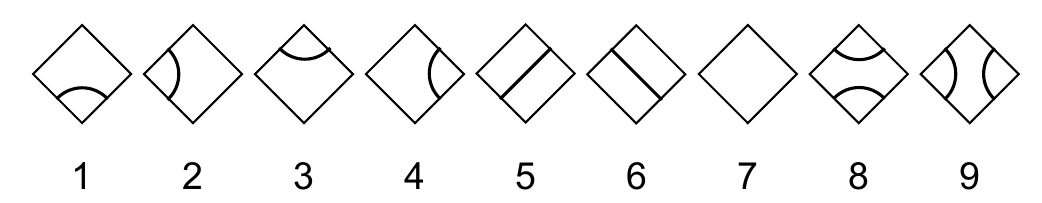}
\caption{The operators acting on the link patterns.}
\label{figplaq}
\end{figure}

Now we define the set of operators which act in the space lp$_L$ of the link patterns 
of length $L$.  These operators are linear combinations of nine basis elements, which 
can be represented graphically by the plaquettes in Fig.  \ref{figplaq}.

\begin{figure}[htb]
\centering
\includegraphics[width=0.7\textwidth]{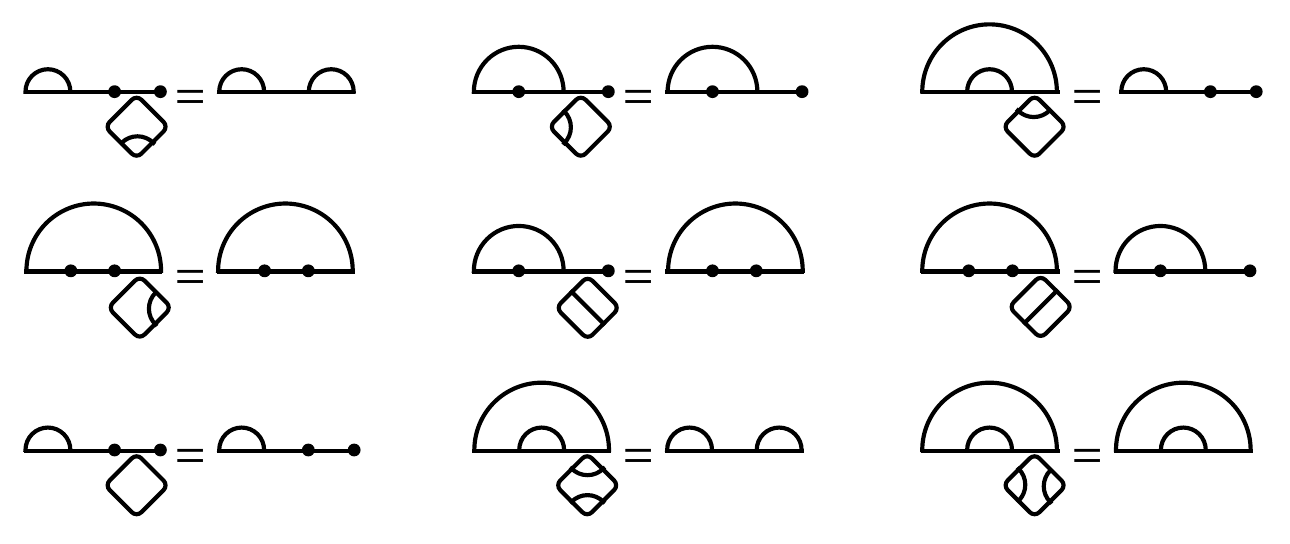}
\caption{Here we present few link patterns on which $\rho^{(i)}$ operators act.}
\label{figaction}
\end{figure}

We  denote these basis operators by $\rho^{(1)}, \rho^{(2)},..,\rho^{(9)}$ in the order of Fig. \ref{figplaq}. 
By $\rho^{(i)}_j$ we denote the operator $\rho^{(i)}$ acting on two 
adjacent sites $j$ and $j+1$ of a link pattern. It is better to describe graphically 
the action of these operators Fig.  \ref{figaction}.

\begin{figure}[htb]
\centering
\includegraphics[width=0.8\textwidth]{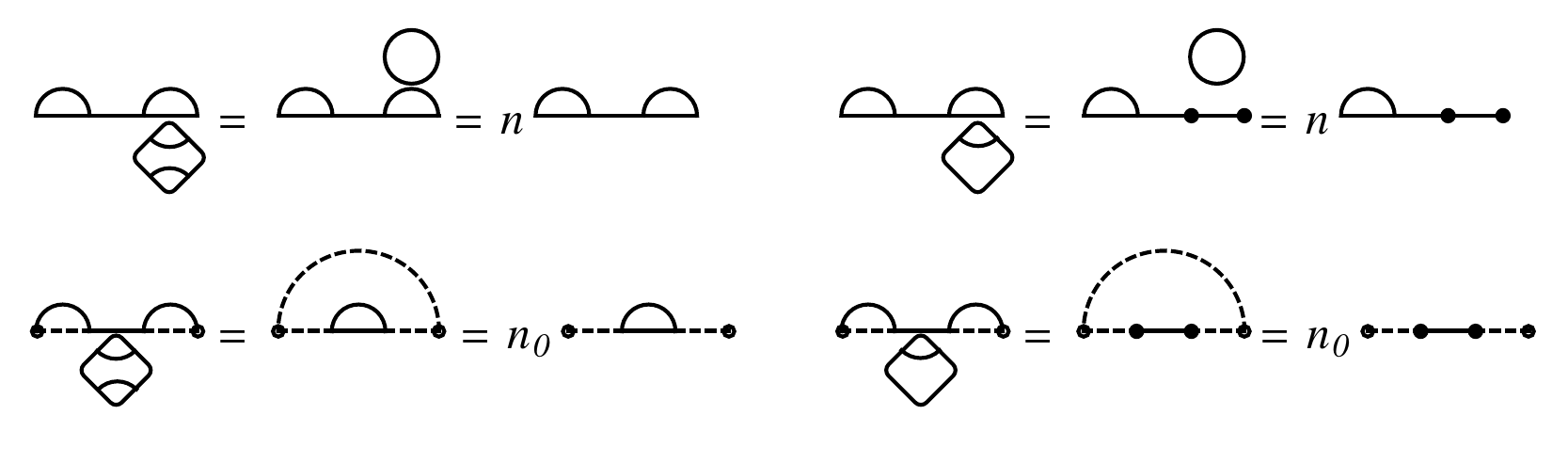}
\caption{The operators $\rho^{(3)}$ and $\rho^{(8)}$ produce a closed loop 
or a line connecting two vertical boundaries, the line is represented by a dashed 
semi-circle.}
\label{figaction1}
\end{figure}

The action of $\rho^{(i)}$ on the link patterns should respect the occupation, i.e., the 
top left and top right occupations of a $\rho^{(i)}_{j}$ should coincide with the 
occupations of the edges $j$ and $j+1$ (respectively) in the link pattern, otherwise the action 
gives zero. We notice that $\rho^{(2)}, \rho^{(4)}, \rho^{(7)}$ and $\rho^{(9)}$ act as projectors on the different occupancies of two edges, $\rho^{(5)}$ and $\rho^{(6)}$ act locally by interchanging an occupied edge 
with an unoccupied edge, the operator $\rho^{(1)}$ also acts locally by inserting a little arch 
at the position $j,~j+1$ of a link pattern.  The two remaining operators 
$\rho^{(3)}$ and $\rho^{(8)}$, 
change the global picture, because they connect two existing paths, or close one.  In this way, $\rho^{(3)}$ and $\rho^{(8)}$ may produce a loop or 
a boundary to boundary link as in Fig.   \ref{figaction1}. We can erase 
the closed loops at the cost of its weight $n$ and the boundary to boundary link 
at the cost of its weight $n_1$.

\begin{figure}[htb]
\centering
\includegraphics[width=1\textwidth]{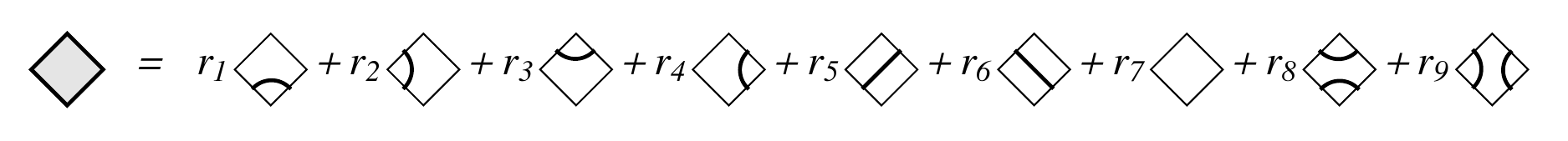}
\caption{$\check{R}$-matrix, represented by the shaded lozenge, can be written as a linear combination of the nine plaquettes given in Fig. \ref{figplaq}.}
\label{figR}
\end{figure}

\begin{figure}[htb]
\centering
\includegraphics[width=0.3\textwidth]{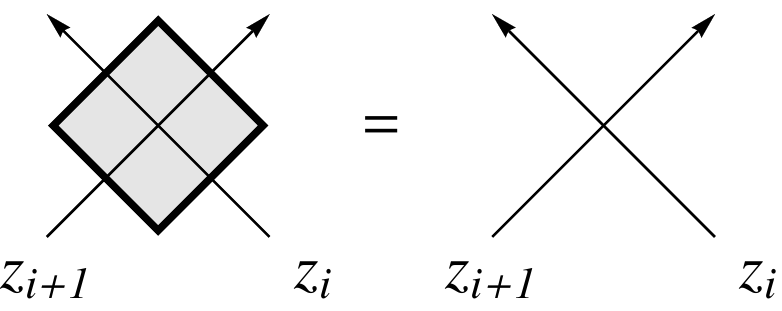}
\caption{The matrix $\check{R}_i(z_i,z_{i+1})$.}
\label{figR2}
\end{figure}

At this point we can turn to the integrability. In order to do this we introduce the 
$\check{R}$-matrix\footnote{We will not use the matrix representations of the operators $R$ and $K$, however, we will still call them the $R$-matrix and the $K$-matrices.} which is a weighted sum of all nine operators 
$\rho_j^{(i)}$ (see Fig. \ref{figR})
\begin{align}\label{Rmx}
\check{R}_j(z_j,z_{j+1})=\sum_{i=1}^{9}\rho_j^{(i)}r_i(z_j,z_{j+1}).
\end{align}
The matrix $\check{R}_j(z_j,z_{j+1})$ acts non trivially on the $j$-th and $j+1$-st spaces 
which carry the rapidities $z_j$ and $z_{j+1}$ respectively, on the rest of the lattice 
spaces it acts as identity. In the graphical notation the rapidities 
are carried by the oriented straight lines as in Fig.  \ref{figR2}.

\begin{figure}[htb]
\centering
\includegraphics[width=0.3\textwidth]{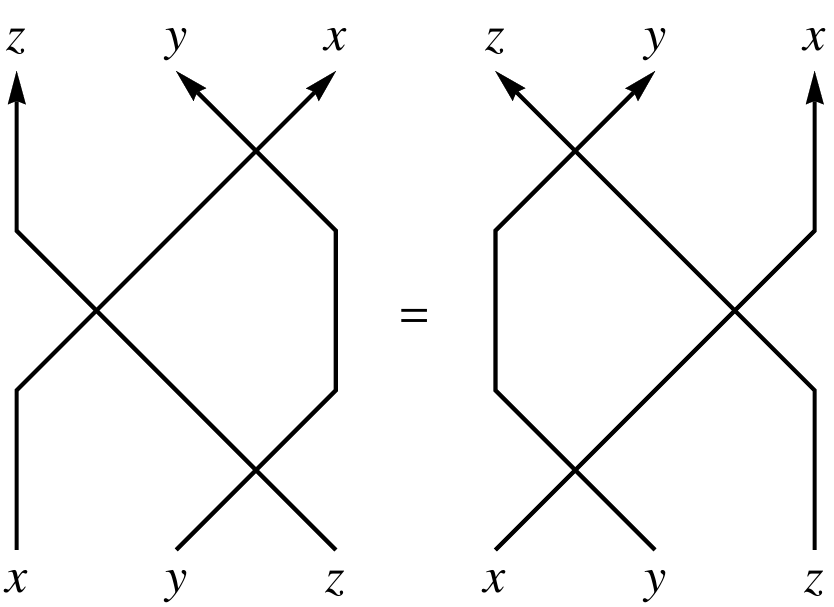}
\caption{The Yang--Baxter equation.}
\label{figYB}
\end{figure}

The integrable dilute O$(n)$ loop model is defined by the $\check{R}$-matrix that satisfies 
the Yang--Baxter (YB) equation \cite{Baxter}.  The $\check{R}$-matrix, in fact, depends only on the ratio of the two spectral parameters, hence we may write 
$\check{R}_j(z_{j}/z_{j+1})\propto \check{R}_j(z_j,z_{j+1})$, then the YB equation reads:
\begin{align}\label{YB}
\check{R}_{i+1}(z/y)\check{R}_{i}(z/x)\check{R}_{i+1}(y/x)=
\check{R}_{i}(y/x)\check{R}_{i+1}(z/x)\check{R}_{i}(z/y).
\end{align}
Graphically it is shown in Fig.   \ref{figYB}.
This equation gives the constraints on the weights $r_i(z)$.  The integrable $R$-matrix
was obtained in \cite{Nienhuis1,Nienhuis2}. In the, so called, additive notation the weights are:
\begin{align}\label{Rweightsad}
&r_1(x)=r_3(x)=\sin 2\lambda \sin x,~~~r_2(x)=r_4(x)=
\sin 2\lambda \sin(3\lambda -x),\nonumber \\
&r_5(x)=r_6(x)=\sin x \sin(3 \lambda -x),~~~r_7(x)=\sin x \sin(3\lambda -x)+
\sin 2\lambda \sin 3\lambda \nonumber \\
&r_8(x)=\sin x \sin (x-\lambda ),~~~ r_9(x)=\sin (x-2\lambda )\sin(x-3\lambda),
\end{align}
and the loop weight is expressed through the loop fugacity $\lambda$: 
\begin{align}\label{loop}
n=-2 \cos 4 \lambda . 
\end{align}
For our purposes the multiplicative notation is more appropriate (already implied 
above). In this case we set: $z=e^{i x}$ and $q=e^{i \lambda}$ in (\ref{Rweightsad}) 
and by abuse of notation we call the resulting weights $r_i(z)$. Up to a common factor of $1/(4 z^2 q^5)$ we have:
\begin{align}\label{Rweights}
&r_1(z)=r_3(z)=-q^3 \left(q^4-1\right) z \left(z^2-1\right),~~~r_2(z)=r_4(z)=
-\left(q^4-1\right) z \left(q^6-z^2\right),\nonumber \\
&r_5(z)=r_6(z)=-q^2 (z^2-1) \left(q^6-z^2\right),~~~r_7(z)=q^8+q^2 z^4-\left(q^2+1\right) \left(q^8-q^4+1\right) z^2, \nonumber \\
&r_8(z)=q^4 \left(z^2-1\right) \left(q^2-z^2\right),~~~ r_9(z)=-\left(q^4-z^2\right) \left(q^6-z^2\right).
\end{align}
The loop weight becomes $n=-q^4-q^{-4}$.
Another operator that we will use is the matrix $R_{i}(z)$. It is simply the 
$\check{R}$ tilted by 45 degrees.

\subsection{Open boundary conditions}
The boundary of the half infinite strip are represented by the two half infinite vertical lines. 
One may consider the periodic boundary conditions, i.e. when the two vertical boundaries 
are identified, closed boundary conditions, i.e. when the loops are reflected from the vertical 
boundaries. 
For general non periodic boundary conditions the loops can end at both 
vertical boundaries. One can also consider open boundary conditions with certain restrictions 
and various mixed boundary conditions.  This is better discussed in \cite{Gt}.
In the case of general open boundary conditions the left and the right 
boundaries carry the boundary spectral parameters. In the conclusion we will mention how to 
recover the ground states corresponding to the other non periodic boundary conditions 
from the ground state corresponding to the general open boundary conditions considered 
here. 

\begin{figure}[htb]
\centering
\includegraphics[width=0.45\textwidth]{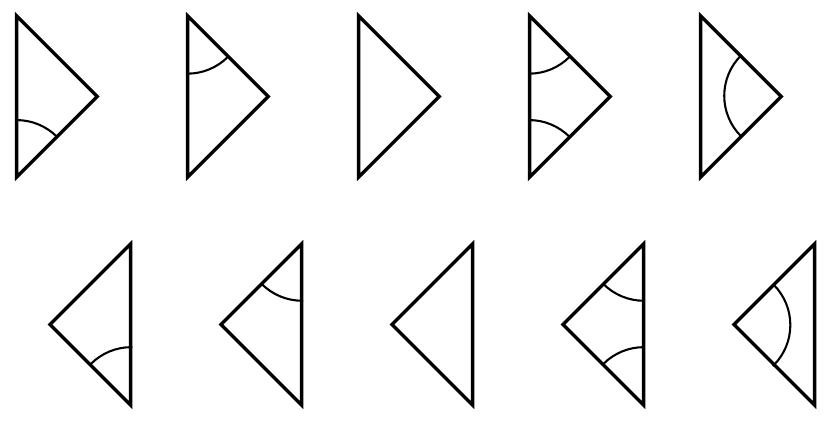}
\caption{The left and the right boundary operators.}
\label{figbplaq}
\end{figure}

\begin{figure}[htb]
\centering
\includegraphics[width=1\textwidth]{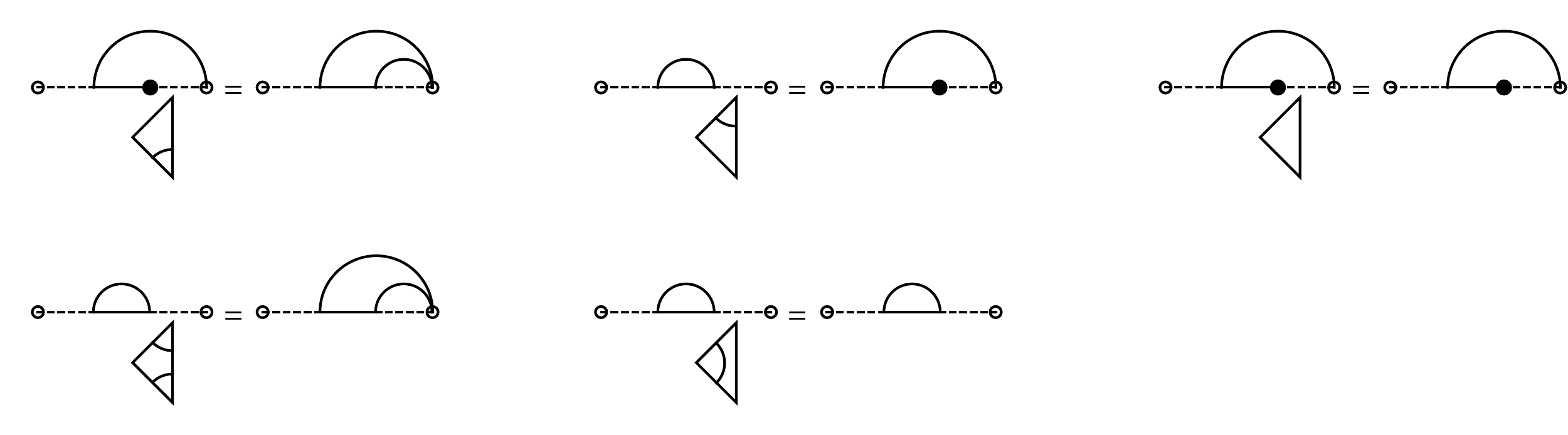}
\caption{The action of the $\kappa_r$ operators.}
\label{figbaction}
\end{figure}

Now we introduce the boundary operators.  These 
operators act on the leftmost and the rightmost points of the link patterns. 
Graphically they are defined by the plaquettes in Fig.  \ref{figbplaq}.
We denote the left boundary operators by $\kappa^{(1)}_l,..\kappa^{(5)}_l$ and the 
right boundary operators by $\kappa^{(1)}_r,..,\kappa^{(5)}_r$, ordered as in Fig. (\ref{figbplaq}). Few examples of their 
action on link patterns are presented in Fig.  \ref{figbaction}.

\begin{figure}[htb]
\centering
\includegraphics[width=0.7\textwidth]{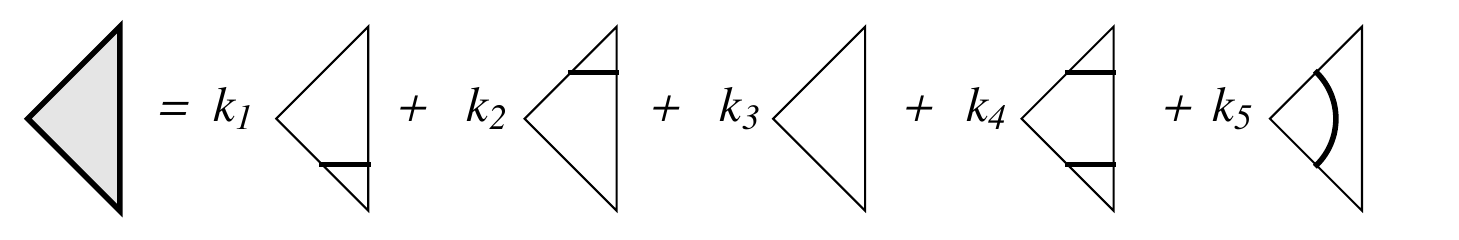}
\caption{The right $K$-matrix is represented by the shaded triangle.}
\label{figKmx}
\end{figure}

We introduce the $K_l$-matrix for the left boundary and the $K_r$-matrix for the right 
boundary.  The $K$-matrices represent the weighted action of the $\kappa$ operators with 
the weights:
$k_{i}(z_1,\zeta_l)$ for the left $K$-matrix and $k_{i}(z_L,\zeta_r)$ for the right 
$K$-matrix. Here, $\zeta_l$ and $\zeta_r$ are the left 
and the right boundary rapidities, therefore (Fig. \ref{figKmx})
\begin{align}\label{Kmx}
K_l(z_1,\zeta_l)=\sum_{i=1}^{5}\kappa^{(i)}_l k_i(z_1,\zeta_l),~~~
K_r(z_L,\zeta_r)=\sum_{i=1}^{5}\kappa^{(i)}_r k_i(z_L,\zeta_r).
\end{align}

\begin{figure}[htb]
\centering
\includegraphics[width=0.4\textwidth]{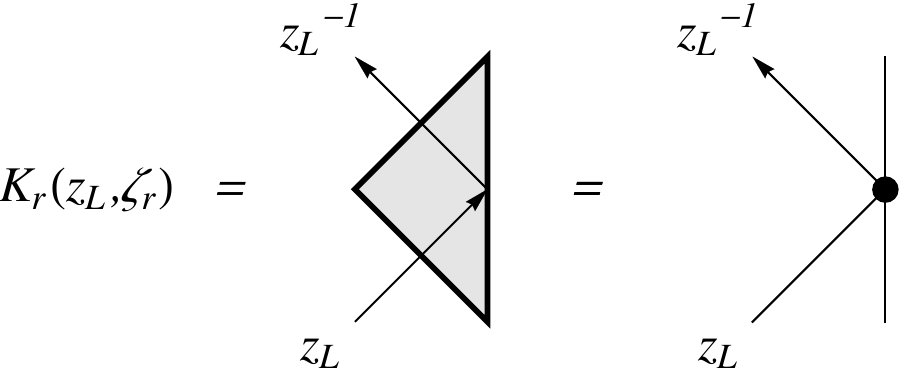}
\caption{The matrix $K_r(z_L,\zeta_r)$.}
\label{figKmx2}
\end{figure}

\begin{figure}[htb]
\centering
\includegraphics[width=0.3\textwidth]{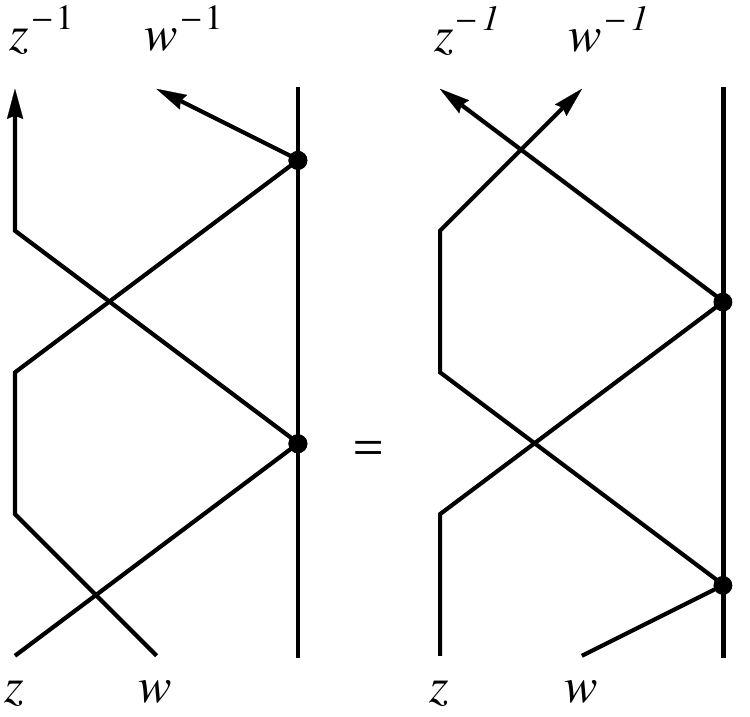}
\caption{The boundary Yang--Baxter equation.}
\label{figBYB}
\end{figure}

The action of $K_i(z,\zeta)$ switches the rapidity from $z$ to $1/z$.  The corresponding 
graphical notation for $K_r(z_{L},\zeta_r)$ is shown in Fig.  \ref{figKmx2}. 
In order to preserve the integrability we need to impose certain conditions on 
the $k$'s, namely the boundary Yang--Baxter (BYB) equation$~$\cite{Skl}. For the right boundary 
it is shown graphically in Fig.   \ref{figBYB} and reads:
\begin{align}\label{BYB}
\check{R}_{L-1}(\frac{w}{z})K_{r}(z,\zeta_r)\check{R}_{L-1}(\frac{1}{w z})K_{r}(w,\zeta_r)=
K_r(w,\zeta_r)\check{R}_{L-1}(\frac{1}{ w z})K_{r}(z,\zeta_r)\check{R}_{L-1}(\frac{w}{z}).
\end{align}
The integrable boundary weights for the right boundary $K$-matrix in the additive convention 
are the following \cite{DJS,dGLR}:
\begin{align}\label{Kweights}
&k_1(x,\zeta)=k_2(x,\zeta)=\zeta \sin 2\lambda \sin 2 x,\nonumber \\
&k_3(x,\zeta)=2 \cos \lambda \sin (\frac{3}{2} \lambda +x)-\zeta^2 n_1 \sin (\frac{1}{2} \lambda +x)
\sin(\frac{1}{2} \lambda -x)\sin(\frac{3}{2} \lambda -x),\nonumber \\
&k_4(x,\zeta)=-\zeta^2 \sin 2 \lambda \sin 2 x \sin(\frac{1}{2} \lambda -x), \nonumber \\
&k_5(x,\zeta)= \sin(\frac{3}{2} \lambda-x) ( 2 \cos \lambda - \zeta^2 n_1 \sin^2(\frac{1}{2}\lambda-x)),
\end{align}
here $n_1$ is the weight of a loop ending on two boundary points. In order to satisfy BYB we need to assume $n_1=n$. There is 
a different solution to the reflection equation, for more details see \cite{dGLR}.

\begin{figure}[htb]
\centering
\includegraphics[width=0.7\textwidth]{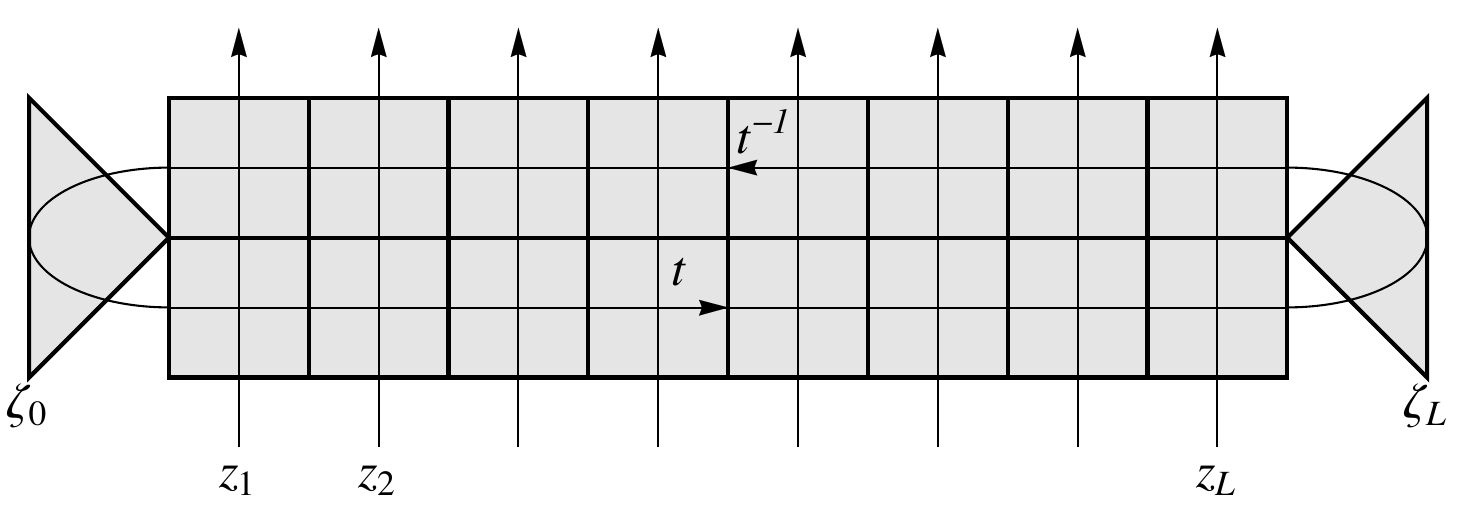}
\caption{The graphical representation of the transfer matrix.}
\label{figTmx}
\end{figure}

In the multiplicative convention: $z=e^{i x}$ and $q=e^{i \lambda}$ in 
 (\ref{Kweights}), the weights read:
\begin{align}\label{Kweightsm}
&k_1(z,\zeta)=k_2(z,\zeta)=2 \zeta  q \left(q^4-1\right) z \left(z^4-1\right),
\nonumber \\
&k_3(z,\zeta)=i q^{1/2} \left(\zeta ^2 n_1 \left(q-z^2\right) \left(q z^2-1\right) \left(q^3
   z^2-1\right)+4 \left(q^2+1\right) z^2 \left(q^3-z^2\right)\right),\nonumber \\
&k_4(z,\zeta)=-i \zeta ^2 q^{1/2} \left(q^4-1\right) \left(z^4-1\right) \left(q z^2-1\right),
 \nonumber \\
&k_5(z,\zeta)= i q^{1/2} \left(q^3 z^2-1\right) \left(\zeta ^2 n_1 \left(q z^2-1\right)^2+4
   \left(q^2+1\right) z^2\right),
\end{align}
where we omitted the denominator $8 z^3 q^3$ with respect to the definitions (\ref{Kweights}). 
The transfer matrix $T(t|z_1,..,z_L;\zeta_l,\zeta_r)$, depicted in Fig.   \ref{figTmx}, 
acts in the space of link patterns. It is constructed from the $R$ and $K$
matrices:
\begin{align}\label{Tmx}
T(t|z_1,..,z_L;\zeta_l,\zeta_r)=\text{Tr}\big{(}
R_1(z_1/t)..R_L(z_L/t)K_r(t,\zeta_r)R_L(1/(t z_L))..R_1(1/(t z_1))K_l(t^{-1},\zeta_l)\big{)},
\end{align}
where the trace means that the lower edge of the $K_l(t^{-1},\zeta_l)$ needs to be identified 
with the left edge of $R_1(t,z_1)$.
Due to the YB and the BYB two transfer matrices with different values of $t$ commute:
\begin{align}\label{TT}
[T(t_1),T(t_2)]=0.
\end{align}

Other important properties of the transfer matrix coming from the YB and BYB are the 
commutations with the $\check{R}$:
\begin{align}\label{RT}
\check{R}_i (z_i,z_{i+1})T(t|z_1,..,z_i,z_{i+1},..,z_L;\zeta_l,\zeta_r)=
T(t|z_1,..,z_{i+1},z_i,..,z_L;\zeta_l,\zeta_r)\check{R}_i (z_i,z_{i+1}),
\end{align}
and with the $K$-matrices
\begin{align}
&K_l (z_1,\zeta_l)T(t|z_1,z_2,..,z_L;\zeta_l,\zeta_r)=
T(t|1/z_1,z_2,..,z_L;\zeta_l,\zeta_r)K_l (z_1,\zeta_l),\label{KTl}\\
&K_r (z_L,\zeta_r)T(t|z_1,..,z_{L-1},z_L;\zeta_l,\zeta_r)=
T(t|z_1,...z_{L-1},1/z_L;\zeta_l,\zeta_r)K_r (z_L,\zeta_r)\label{KTr}
\end{align}

The common ground-state vector of the family of transfer matrices $T(t|z_1,..,z_L;\zeta_l,\zeta_r)$ 
parametrized by $t$ we denote by $\Psi_L$:
\begin{align}\label{Tmxeq}
T(t|z_1,..,z_L;\zeta_l,\zeta_r)\Psi_L (z_1,..,z_L;\zeta_l,\zeta_r)\propto
\Psi_L (z_1,..,z_L;\zeta_l,\zeta_r).
\end{align}
The eigenvectors of $T$ can be written in the link pattern basis, in particular:
\begin{align}\label{Psieq}
\Psi_L (z_1,..,z_L;\zeta_l,\zeta_r)=\sum_{\pi \in \text{LP}_L} 
\psi_{\pi} (z_1,..,z_L;\zeta_l,\zeta_r)|\pi\rangle.
\end{align}
Our aim is to understand how to compute the components $\psi_{\pi}$ when 
the loop weight $n$ (and also $n_1$) is equal to $1$.

\section{Loop weight \texorpdfstring{$n=1$}{Lg}}\label{sec3}
In this section we specify our model to the case when $q=e^{i\pi/3}$, then the loop 
weight $n$ (and we assume also $n_1=n$) has the weight equal 
to$~1$.  This means that the presence of the closed loops in a configuration does not affect 
the weight of this configuration. At this point the ground state is a steady 
state of a stochastic process that can be defined by the Hamiltonian (or transfer matrix) of the 
dTL model. This happens in analogy with the dense TL loop model at $n=1$ \cite{PRGN}. 
 The ground state entries become the relative 
probabilities of the occurrence of the corresponding link patterns. Assuming that the 
transfer matrix is normalized, the ground state eigenvector equation becomes:
\begin{align}\label{Tmxeq1}
T(t|z_1,..,z_L;\zeta_l,\zeta_r)\Psi_L (z_1,..,z_L;\zeta_l,\zeta_r)=
\Psi_L (z_1,..,z_L;\zeta_l,\zeta_r).
\end{align}
This implies the invariance of the probabilities under the addition of two rows of the transfer 
matrix.  Since the transfer matrix is a rational function of the rapidities, we 
can normalize $\Psi_L$ such that its components become coprime polynomials. 

We will replace everywhere $q$ by $\omega$ and assume $\omega=e^{i\pi/3}$.  The matrices $R$ and 
$K$ simplify. The $R$-matrix given in (\ref{Rweights}) has a common factor of $\omega/(1-z^2)$ when $q=\omega$, we omit this factor below and use $r_i(z)$ to denote the weights. 
\begin{align}\label{Rweights1}
&r_1(z)=r_2(z)=r_3(z)=r_4(z)=\omega  (\omega +1) z,~~~
r_5(z)=r_6(z)=r_7(z)=z^2-1, \nonumber \\
&r_8(z)=-\left(\omega +z\right) \left(\omega ^2 z+1 \right),
~~~ r_9(z)=\left(\omega ^2+z\right) \left(\omega  z+1 \right).
\end{align}
From now on we will use the parameter $\tilde{\zeta}=-\omega \zeta/2$ instead of $\zeta$ in what follows and in order to simplify the formulae we will omit the tilde. The left $K$-matrix weights read 
\begin{align}\label{Kweights1}
&k_{1,l}(z,\zeta_l)=k_{2,l}(z,\zeta_l)=-\frac{(\omega+1) \left(z^2-1\right)}{z},~~~
k_{3,l}(z,\zeta_l)=\frac{\zeta ^2 \omega+\zeta ^2 \omega z^4-\zeta ^2 \omega z^2+z^2}{\zeta  z^2}, \nonumber \\
&k_{4,l}(z,\zeta_l)=-\frac{\zeta  (\omega+1) \left(z^2-1\right) \left(\omega-z^2\right)}{z^2},
~~~ k_{5,l}(z,\zeta_l)=-\frac{\left(-\zeta  \omega+\zeta  z^2-z\right) \left(-\zeta  \omega+\zeta  z^2+z\right)}{\zeta 
   z^2}.
\end{align}
The weights of the right boundary $K$-matrix are given by $k_{i,r}(z,\zeta)=k_{i,l}(1/z,\zeta)$., they agree with (\ref{Kweightsm}) up to the denominator $4 \zeta z^2 (z^2 + 1)$.

\begin{figure}[htb]
\centering
\includegraphics[width=0.6\textwidth]{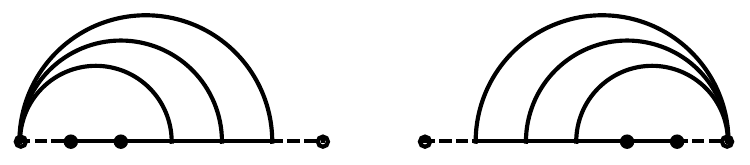}
\caption{Two link patterns $\pi_2$ and $\tilde{\pi}_2$ at $L=5$.}
\label{fignested}
\end{figure}

\begin{figure}[htb]
\centering
\includegraphics[width=0.6\textwidth]{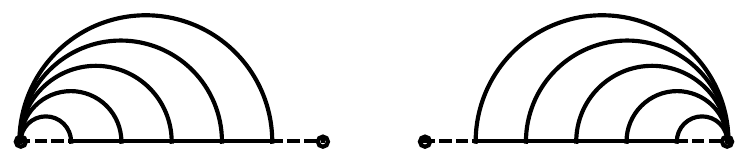}
\caption{Two $L=5$ link patterns $\pi_0$ and $\tilde{\pi}_0$.}
\label{figfn}
\end{figure}

\begin{figure}[htb]
\centering
\includegraphics[width=0.8\textwidth]{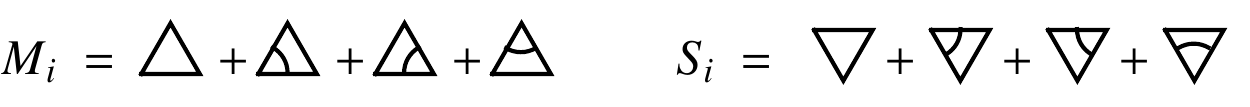}
\caption{The operators $M_i$ and $S_i$.}
\label{figMS}
\end{figure}

\begin{figure}[htb]
\centering
\includegraphics[width=0.5\textwidth]{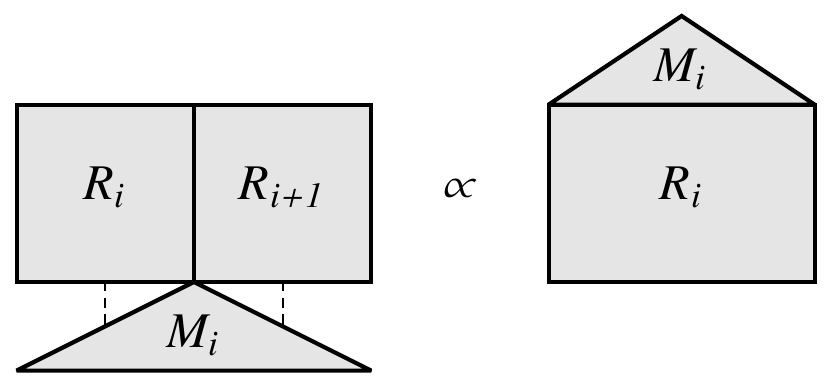}
\caption{Equation  (\ref{MRR}).}
\label{figYenB}
\end{figure}

Acting with the $\check{R}$-matrix on a link pattern gives a sum of the link 
patterns obtained from the action of the $\rho^{(j)}$ operators, weighted by the corresponding functions $r_j$. For example, on a link pattern with two empty sites $n_i=0,~n_{i+1}=0$, 
$\check{R}_i$ acts by 
$\rho^{(1)}$ with the probability $r_1$ and by $\rho^{(7)}$ with the probability $r_7$. 
Hence, we need to normalize this action with $W=r_1+r_7$. In fact, the weights 
 (\ref{Rweights1}) 
are such that for any occupation $n_i$ and $n_{i+1}$ the normalization is 
the same. 
\begin{align}\label{Rnorm}
W(z_i,z_{i+1})=\left(\omega  z_i+z_{i+1}\right) \left(\omega ^2 z_i+z_{i+1}\right).
\end{align}
Similar arguments apply to the $K$-matrices.  The normalizations of $K_l$ and $K_r$ are: 
\begin{align}
U_l(z,\zeta_l)=-\frac{\left(\omega \zeta _l-z^2 \zeta _l+z\right)
 \left(-\zeta _l+\omega z^2 \zeta   _l-z\right)}{z^2 \zeta _l}, \label{Knorml}\\
U_r(z,\zeta_r)=\frac{\left(-\omega  \zeta _r+z^2 \zeta _r+z\right)
 \left(-\zeta _r+\omega  z^2 \zeta   _r+z\right)}{z^2 \zeta _r}.\label{Knormr}
\end{align}
Now, using (\ref{RT}) we find
\begin{align}\label{RqKZ}
\check{R}(z_i,z_{i+1})\Psi_L (z_1,..,z_i,z_{i+1},..,z_L;\zeta_l,\zeta_r)=
W(z_i,z_{i+1})\Psi_L (z_1,..,z_{i+1},z_{i},..,z_L;\zeta_l,\zeta_r),
\end{align}
and also, using  (\ref{KTl}) and  (\ref{KTr}):
\begin{align}
&K_l(z_1,\zeta_l)\Psi_L (z_1,..,z_L;\zeta_l,\zeta_r)=
U_l(z_1,\zeta_l)\Psi_L (1/z_1,..,z_L;\zeta_l,\zeta_r),\label{BqKZl} \\
&K_r(z_L,\zeta_r)\Psi_L (z_1,..,z_L;\zeta_l,\zeta_r)=
U_r(z_L,\zeta_r)\Psi_L (z_1,..,1/z_L;\zeta_l,\zeta_r).\label{BqKZr}
\end{align}
Equations  (\ref{RqKZ})- (\ref{BqKZr}) together will be called the $q$KZ equations.  This set of 
equations is a system of functional equations with polynomial solutions. 
It allows to find all 
components of the ground state only if a certain set of the components is already known. 
These components $\psi_{\pi}$ are those which correspond to the link patterns 
$\pi_i=\{0_1,..,0_{i},-1_{i+1},..,-1_L\}$ or  
$\tilde{\pi}_i=\{1_1,..,1_{L-i-1},0_{L-i},..,0\}$
(here, the index denotes the corresponding site).  These are the components which have 
all empty sites starting from the first (respectively, last) site up to the $i$-th 
($L-i$-th) and the rest is connected to the left (right) boundary (Fig. \ref{fignested}). 
The elements $\psi_{\pi_0}$ and $\psi_{\tilde{\pi}_0}$ are called the fully nested elements 
(Fig. \ref{figfn}).  They play a special role in our computations since all $\psi_{\pi_i}$ 
as well as $\psi_{\tilde{\pi}_i}$ can be obtained from the 
fully nested elements of larger systems using the recurrence relation to which we 
turn or discussion now.
We notice that when we set two consecutive rapidities $z_i$ and $z_{i+1}$ to 
$z_i \omega$ and $z_{i}/\omega$ the matrix $\check{R}(z_i,z_{i+1})$ factorizes into two 
operators
\begin{align}\label{RMS}
\check{R_i}(z\omega,z/\omega)=(\omega^2+\omega)z^2 S_i M_i.
\end{align}
Now using the operator $M_i$ we can write 
\begin{align}\label{MRR}
&M_i\check{R_i}(z_i \omega)R_{i+1}(z_i/\omega) =(z_i^2-1) R_i(t,z_i)M_i,\nonumber \\
&M_i\check{R_i}(z_i/\omega)R_{i+1}(z_i \omega) =(z_i^2-1) R_i(t,z_i)M_i,
\end{align}
which can be checked by a direct calculation. 
This equation means that $M$ maps two sites into one site and hence merges the 
two $R$-matrices into one after the substitution $z_i=z_i \omega$ and 
$z_{i+1}=z_i/\omega$.  
The graphical representations of $M$, $S$ and  (\ref{MRR}) are presented in the
figures  (\ref{figMS}) and  (\ref{figYenB}).

One can see from (\ref{MRR}) (Fig.  \ref{figYenB}) and the definition of the transfer matrix (\ref{Tmx}) that by applying $M_i$ to the transfer matrix we get:
\begin{align}\label{MT}
M_i T_{L+1}(t|z_1,..,z_i \omega,z_i/\omega,z_{i+1},..,z_L;\zeta_l,\zeta_r)\propto
T_L(t|z_1,..,z_i,z_{i+1},..,z_L;\zeta_l,\zeta_r)M_i,
\end{align}
where we included the indices $L$ and $L+1$ in the transfer matrix to denote the length of the 
space on which it acts. The proportionality factor is coming from the factors on the right hand sides of (\ref{MRR}) and can be eliminated by choosing an appropriate normalization of the transfer matrix. Applying  (\ref{MT}) to the ground state we get:
\begin{align}\label{MGS0}
&M_i T_{L+1}(..,z_i \omega,z_i/\omega,z_{i+1},..)
\Psi_{L+1}(..,z_i \omega,z_i/\omega,z_{i+1},..)\propto\nonumber  \\
&T_L(..,z_i,z_{i+1},..)M_i
\Psi_{L+1}(..,z_i \omega,z_i/\omega,z_{i+1},..),
\end{align}
which becomes
\begin{align}\label{MGS1}
M_i \Psi_{L+1}(..,z_i \omega,z_i/\omega,z_{i+1},..)=
T_L(t|..,z_i,z_{i+1},..)M_i
\Psi_{L+1}(..,z_i \omega,z_i/\omega,z_{i+1},..),
\end{align}
where we  omit the dependence on the irrelevant variables for convenience. We obtain 
the following recurrence relation:
\begin{align}\label{MGS}
M_i \Psi_{L+1}(z_1,..,z_i \omega,z_i/\omega,z_{i+1},..,z_{L+1};\zeta_l,\zeta_r)\propto
\Psi_L(z_1,..,z_i,z_{i+1},..,z_L;\zeta_l,\zeta_r).
\end{align}
Let us introduce the operator $\mu_i$ which projects the ring of polynomials of $L$ variables $\mathbb{C}[z_1,..,z_L ]$ onto the ring of polynomials of $L-1$ variables $\mathbb{C}[z_1,..,\hat{z}_{i+1},..z_{L-1}]$, where $\hat{z}_{i+1}$ means that this variable is absent from the list. Consider a polynomial $f(z_1,..,z_L)\in \mathbb{C}[z_1,..,z_L]$, the action of $\mu_i$ is defined by
\begin{align}\label{mu}
\mu_i f(z_1,..,z_{i-1},z_{i},z_{i+1},z_{i+2},..,z_L)=
f(z_1,..,z_{i-1},z_i \omega,z_i/\omega,z_{i+2},..,z_{L}).
\end{align}
For example, fix $L=3$ and set $f(z_1,z_2,z_3)=z_1^2+ z_2 z_3+ z_3^2$, then the action of $\mu_1$ and $\mu_2$ is 
\begin{align*}
&\mu_1 \left( z_1^2+ z_2 z_3+ z_3^2 \right) =  \left(z_1 \omega \right)^2+ \left(z_1/\omega\right) z_3+ z_3^2, \\
&\mu_2 \left( z_1^2+ z_2 z_3+ z_3^2 \right) = z_1^2+ \left(z_2\omega\right) \left(z_2/\omega\right)+ \left(z_2/\omega \right)^2.
\end{align*}
The recurrence  (\ref{MGS}) relates the components of $\Psi_{L+1}$ to the components of 
$\Psi_L$. What we need to do to complete this equation is to find the proportionality factor. 
In the next section we will see that the knowledge of the fully nested element 
allows one to find this proportionality factor. Combining the boundary $q$KZ equation 
with  (\ref{MGS}) will allow us to find all $\psi_{\pi_i}$.  Then we will show 
how to recover the full ground state using the $q$KZ system.

\subsection{The \texorpdfstring{$q$}{Lg}KZ equations and the fully nested elements}

Let us take a closer look at the $q$KZ equations. First we focus on the 
bulk $q$KZ. (\ref{RqKZ}) gives $(L-1)$ equation for each link pattern, hence 
$(L-1)3^L$ equations in total. It is sufficient to write a few distinct cases 
depending 
on the local connectivity at the sites $i$ and $i+1$ of a link pattern $\pi$ of the 
component $\psi_{\pi}$. For convenience, we will write $\pi=\{\alpha,n_i,n_{i+1},\beta\}$, 
where $\alpha$ and $\beta$ are the parts of $\pi$ on the left and on the right 
to the sites $n_i$ and $n_{i+1}$, respectively. 
Also, since we will focus only on the sites $i$ and $i+1$ we will 
not write explicitly the dependence on the variables attached to the other spaces.

1. Both sites $i$ and $i+1$ are occupied.

$~~~$a.) The sites $i$ and $i+1$ in $\pi$ are connected via a little arch: 
$\pi=\{\alpha,1,-1,\beta\}$. In this case  (\ref{RqKZ}) gives:
\begin{align}\label{qKZarch}
&W(z_i,z_{i+1})\psi_{\pi}(..,z_{i+1},z_i,..)=
r_9(z_i,z_{i+1})\psi_{\pi}(..,z_{i},z_{i+1},..)+\nonumber \\
&r_1(z_i,z_{i+1})\psi_{\{\alpha,0,0,\beta\}}(..,z_i,z_{i+1},..)
+r_8(z_i,z_{i+1})\sum_{\pi^{\prime}: \rho_i^{(8)} \pi^{\prime}=\pi}
\psi_{\pi^{\prime}}(..,z_i,z_{i+1},..).
\end{align}

$~~~$b.) The sites $i$ and $i+1$ are not connected to each other:
\begin{align}\label{qKZv2}
W(z_i,z_{i+1})\psi_{\pi}(..,z_{i+1},z_i,..)=r_9(z_i,z_{i+1})\psi_{\pi}(..,z_{i},z_{i+1},..).
\end{align}

2. Both sites $i$ and $i+1$ in $\pi$ are unoccupied
\begin{align}\label{qKZem}
&W(z_i,z_{i+1})\psi_{\pi}(..,z_{i+1},z_i,..)=
r_7(z_i,z_{i+1})\psi_{\pi}(..,z_{i},z_{i+1},..)+ \nonumber \\
&r_3(z_i,z_{i+1})\sum_{\pi^{\prime}: \rho_i^{(3)} \pi^{\prime}=\pi}
\psi_{\pi^{\prime}}(..,z_i,z_{i+1},..).
\end{align}

3. Finally, one of the sites $i$ and $i+1$ in $\pi$ is occupied and the other one is 
unoccupied.  There are two distinct cases. Fix $i$, then in both cases we will denote by $\pi^{\prime}$ the 
link pattern that is obtained from $\pi$ by simply interchanging the occupations $n_i$
and $n_{i+1}$ of the sites $i$ and $i+1$ in $\pi$. Both equations have the same form:
\begin{align}\label{qKZhop1}
&W(z_i,z_{i+1})\psi_{\pi}(..,z_{i+1},z_i,..)=
r_2(z_i,z_{i+1})\psi_{\pi}(..,z_{i},z_{i+1},..)+
r_5(z_i,z_{i+1})\psi_{\pi^{\prime}}(..,z_i,z_{i+1},..).
\end{align}
This equation corresponds to the case when $i$ is occupied and $i+1$ is empty, 
the other equation, i.e. the one for $i$-empty and $i+1$-occupied, 
is obtained from this one by replacing $r_2$ with $r_4$ and $r_5$ with $r_6$, 
however, in both replacements the corresponding weights are equal.

Now let us turn to the boundary $q$KZ equations. Here, we also have to consider 
few cases separately. Since the logic is similar for both boundaries we will treat 
the left boundary only.

1.  The first site is occupied.

$~~~$a.) The connectivity at the first site is $\pi=\{1,\beta\}$, that means 
that this site is connected to another site or the boundary on the right:
\begin{align}\label{bqKZ1}
&U_l(z_1,\zeta_l)\psi_{\pi}(1/z_1,..)=
k_{l,5}(z_1,\zeta_l)\psi_{\pi}(z_{1},..).
\end{align}

$~~~$b.) The connectivity at the first site is $\pi=\{-1,\beta\}$, which means it is connected to 
the left boundary:
\begin{align}\label{bqKZ2}
&U_l(z_1,\zeta_l)\psi_{\pi}(1/z_1,..)=
k_{l,5}(z_1,\zeta_l)\psi_{\pi}(z_{1},..)+k_{l,1}\psi_{\{0,\beta\}}(z_{1},..)+\nonumber \\
&k_{l,4}(\psi_{\{1,\beta\}}(z_{1},..)+\psi_{\pi}(z_{1},..)),
\end{align}

2.  The site $1$ is unoccupied $\pi=\{0,\beta\}$:
\begin{align}\label{bqKZ3}
&U_l(z_1,\zeta_l)\psi_{\pi}(1/z_1,..)=
k_{l,3}(z_1,\zeta_l)\psi_{\pi}(z_{1},..)+k_{l,2}(\psi_{\{1,\beta\}}(z_{1},..)
+\psi_{\{-1,\beta\}}(z_{1},..)),
\end{align}

As we already mentioned, finding the fully nested elements is the first important 
step. Let us take a look at the $q$KZ equations which involve the fully nested elements, 
i.e. (\ref{qKZv2}), first for $\pi_0$. Let us fix $i=1$, then this equation 
after canceling the common factors gives:
\begin{align}\label{qKZfn1}
(\omega z_1+z_2)\psi_{\pi_0}(z_2,z_1,..)=
(\omega z_2+z_1)\psi_{\pi_0}(z_1,z_2,..).
\end{align}
This means $\psi_{\pi_0}$ has the factor $(\omega z_1+z_2)$ and is symmetric in $z_1, z_2$
in the remaining part, which we denote by $\psi^{(1)}_{\pi_0}$: 
\begin{align}\label{qKZfn2}
\psi_{\pi_0}(z_1,z_2,..)=(\omega z_1+z_2)\psi^{(1)}_{\pi_0}(z_1,z_2,..).
\end{align}
Now if we fix $i=2$ in  (\ref{qKZv2}) we see that $\psi^{(1)}_{\pi_0}$ must contain 
the factor $(\omega z_2+z_3)$ and taking into account its symmetries it also must contain 
the factor $(\omega z_1+z_3)$, we get 
\begin{align}\label{qKZfn3}
\psi_{\pi_0}(z_1,z_2,..)=(\omega z_1+z_2)(\omega z_2+z_3)(\omega z_1+z_3)
\psi^{(2)}_{\pi_0}(z_1,z_2,z_3,..).
\end{align}
We can proceed in the same manner for $i=3,..,L$ and what we obtain in the end is: 
\begin{align}\label{qKZfnL}
\psi_{\pi_0}(z_1,z_2,..)=\psi^{(L-1)}_{\pi_0}(z_1,..,z_L)\prod_{1\leq i<j\leq L}(\omega z_i+z_j).
\end{align}
Here, $\psi^{(L-1)}_{\pi_0}$ is symmetric in all variables. 

The right boundary $q$KZ equation that involves the $\psi_{\pi_0}$ (the right boundary 
analog of  (\ref{bqKZ1})) gives:
\begin{align}\label{bqKZfn}
\frac{ \left(-\zeta_r +\zeta_r  \omega z_L^2-z_L\right)}{\zeta_r 
   z_L}\psi _{\pi _0}\left(..,1/z_L\right)=
-\frac{\left(-\zeta_r  \omega+\zeta_r  z_L^2+z_L\right)}{\zeta_r  z_L}\psi _{\pi _0}(..,z_L),
\end{align}
which holds when:
\begin{align}\label{bqKZfn1}
\psi _{\pi _0}(..,z_L)=\frac{ \left(-\zeta_r +\omega \zeta_r  z_L^2-z_L\right)}{\zeta_r 
   z_L}\bar{\psi} _{\pi _0}(..,z_L),
\end{align}
and $\bar{\psi} _{\pi _0}(..,z_L)=\bar{\psi} _{\pi _0}(..,1/z_L)$. At this point it is 
convenient to reparametrize $\zeta_r$ and $\zeta_l$ as follows: 
\begin{align}\label{zeta}
\zeta_r=\omega \frac{x_r}{x_r^2+1},~~~\zeta_l=\omega \frac{x_l}{x_l^2+1}.
\end{align}
Then, equation  (\ref{bqKZfn1}) reads:
\begin{align}\label{bqKZfn2}
\psi _{\pi _0}(..,z_L)=\frac{\omega (\omega x_r+z_L) (\omega+x_r z_L)}{x_r z_L}
\bar{\psi} _{\pi _0}(..,z_L).
\end{align}
Combining this with (\ref{qKZfnL}) we obtain: 
\begin{align}\label{fnfactor}
\psi_{\pi_0}(z_1,z_2,..,z_L;\zeta(z_{L+1}))= \psi_{\pi_0}^*(z_1,..,z_L,z_{L+1})
\prod_{1\leq i<j\leq L+1}\frac{(\omega z_i+z_j)(1+\omega z_i z_j)}{z_i z_j},
\end{align}
where we replaced $x_r$ by $z_{L+1}$. 
The prefactor in this expression $\psi_{\pi_0}^*$ is symmetric in all $z_i$'s
as well as in $z_i\rightarrow z_i^{-1}$.  That is all what we can deduce from 
the $q$KZ system about the fully nested element $\psi_{\pi_0}$. In the same manner 
we find that: 
\begin{align}\label{fntfactor}
\psi_{\tilde{\pi}_0}(z_1,z_2,..,z_L;\zeta(z_0))= \psi_{\tilde{\pi}_0}^*(z_0,z_1,..,z_L)
\prod_{0\leq i<j\leq L}\frac{(\omega z_i+z_j)(\omega+ z_i z_j)}{z_i z_j}.
\end{align}
Here $z_{0}$ replaces $x_l$ and the prefactor $\psi_{\tilde{\pi}_0}^*$ is symmetric in 
$\{z_0^{\pm 1},z_1^{\pm 1},..,z_L^{\pm 1}\}$. Both $\psi_{\pi_0}^*$ and 
$\psi_{\tilde{\pi}_0}^*$ are transparent to the $q$KZ equations, hence we assume that 
they do not depend on $z_i$'s. 

\subsection{The recurrence relation}
\begin{figure}[htb]
\centering
\includegraphics[width=0.3\textwidth]{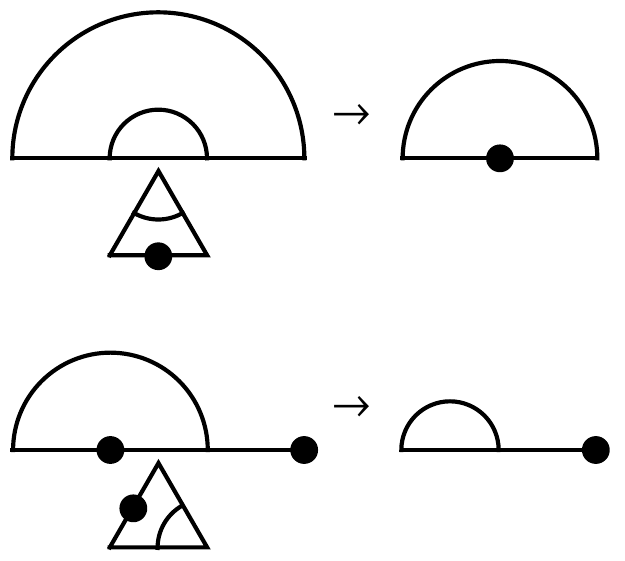}
\caption{The action of the operator $M_2$.}
\label{figMaction}
\end{figure}

In this section we complete  (\ref{MGS}) by deriving the 
proportionality factor. In order to do that let us figure out first what does the 
mapping  (\ref{MGS}) mean for the components of the ground state.  This mapping consists 
of two parts: the action of $\mu_i$ and the action of $M_i$. Since the operator $M_i$ 
maps the link patterns of size $L+1$ to the link patterns of size $L$ it induces 
the correspondence between the components of $\Psi_{L+1}$ and $\Psi_L$. $M_i$ acts 
on the sites $i$ and $i+1$ by mapping the link patterns of LP$_{L+1}$ 
with $n_i=1,~n_{i+1}=-1$ or $n_i=0,~n_{i+1}=0$ to the link patterns of LP$_L$ with 
$n_i=0$, more explicitly $\{\alpha,1,-1,\beta\}\in\text{LP}_{L+1}$ maps to $\{\alpha,0,\beta\}\in\text{LP}_{L}$ and $\{\alpha,0,0,\beta\}\in\text{LP}_{L+1}$ maps to $\{\alpha,0,\beta\}\in\text{LP}_{L}$. 
The link patterns in LP$_{L+1}$ which have 
$n_i=\pm 1,~n_{i+1}=0$ or $n_i=0,~n_{i+1}=\pm 1$ are mapped to the link patterns of LP$_L$ 
with $n_i=\pm 1$, for example $\{\alpha,1,0,\beta\}\in\text{LP}_{L+1}$ maps to $\{\alpha,1,\beta\}\in\text{LP}_{L}$
(Fig. \ref{figMaction}). 
Note, however, that all components 
with $n_i=-1,~n_{i+1}=\pm 1$ and $n_i=1,~n_{i+1}= 1$, due to the $q$KZ equation 
 (\ref{qKZv2}) acquire the factor of $z_i \omega +z_{i+1}$.  This factor vanishes under the 
action of $\mu_i$ since $\omega=e^{i\pi/3}$.  Therefore for some link patters $\alpha$ and $\beta$ the recurrence  (\ref{MGS}) implies
\begin{align}
&\mu_i\psi_{\alpha,n_i=-1,n_{i+1}=\pm 1,\beta}(..,z_i,z_{i+1},..)=
\mu_i\psi_{\alpha,n_i=1,n_{i+1}=1,\beta}(..,z_i,z_{i+1},..)=0, \label{mum1}\\
&\mu_i\psi_{\alpha,n_i=1,n_{i+1}=-1,\beta}(..,z_i,z_{i+1},..)=
F_i(z_1,..,\hat{z}_{i+1},..,z_L)\psi_{\alpha,n_i=0,\beta}(..,z_i,..), \label{mu1m1} \\
&\mu_i\psi_{\alpha,n_i=1,n_{i+1}=0,\beta}(..,z_i,z_{i+1},..)=
F_i(z_1,..,\hat{z}_{i+1},..,z_L)\psi_{\alpha,n_i=1,\beta}(..,z_i,..), \label{mu10}\\
&\mu_i\psi_{\alpha,n_i=0,n_{i+1}=1,\beta}(..,z_i,z_{i+1},..)=
F_i(z_1,..,\hat{z}_{i+1},..,z_L)\psi_{\alpha,n_i=1,\beta}(..,z_i,..),\label{mu01}\\
&\mu_i\psi_{\alpha,n_i=0,n_{i+1}=0,\beta}(..,z_i,z_{i+1},..)=
F_i(z_1,..,\hat{z}_{i+1},..,z_L)\psi_{\alpha,n_i=0,\beta}(..,z_i,..).\label{mu00}
\end{align}
The factor $F_i(z_1,..,z_L)$ is exactly 
the proportionality factor in  (\ref{MGS}), and the label $i$ denotes the site on 
which $\mu_i$ is applied. $F_i$ also depends on $\zeta_l$ and $\zeta_r$ which 
we did not indicate explicitly.

Now we would like to obtain the explicit form of the function $F_i$. In order to do that 
we need to consider the left boundary $q$KZ equation for the element $\psi_{\pi_0}$ 
which is given by  (\ref{bqKZ2}) and apply twice the $\mu_1$ operator to this equation. 
We find
\begin{align}\label{mubqKZ}
&\mu_1 U_l(z_1,\zeta_l)\psi_{\pi_0}(1/z_1,..)=
\mu_1 k_{l,5}(z_1,\zeta_l) \psi_{\pi_0}(z_{1},..)+\nonumber \\
&\mu_1 k_{l,1}(z_1,\zeta_l)\psi_{\pi_1}(z_{1},..)+
\mu_1 k_{l,4}(z_1,\zeta_l)(\psi_{\{1,-1..,-1\}}(z_{1},..)+
 \psi_{\pi_0}(z_{1},..)).
\end{align}
Using  (\ref{mum1}),  (\ref{mu01}) and  (\ref{mu1m1}) we obtain:
\begin{align}\label{mubqKZ2}
&U_l(z_1\omega,\zeta_l)\psi_{\pi_0}(1/(z_1 \omega),z_1/\omega,z_2,..)=\nonumber \\
&F_1(z_1,..,\hat{z}_{i+1},..)(k_{l,1}(z_1\omega,\zeta_l)\psi_{\pi_0}(z_{1},..)+
k_{l,4}(z_1 \omega,\zeta_l)\psi_{\pi_1}(z_1,..)).
\end{align}
Applying $\mu_1$ again we arrive at an equation containing the $F$'s and the fully nested elements. It can be rewritten in the form
\begin{align}\label{mubqKZ3}
& U_l(z_1\omega^2,\zeta_l)\psi_{\pi_0}(1/(z_1 \omega^2),z_1,z_1/\omega,z_2,..,z_L)=\nonumber \\
& F_1(z_1 \omega,z_1/ \omega,..,z_L)F_1(z_1,..,z_L)
k_{l,4}(z_1 \omega^2,\zeta_l)\psi_{\pi_0}(z_1,..,z_L),
\end{align}
where $\psi_{\pi_0}$ appearing on the left hand side depends on $L+2$ arguments and is the $\pi_0$ component of the ground state on the strip with $L+2$ sites while $\psi_{\pi_0}$ on the right and side depends on $L$ arguments and is the $\pi_0$ component of the ground state on the strip with $L$ sites. 
Plugging here the explicit form of $\psi_{\pi_0}$ we find that the choice
\begin{align}\label{factor}
F_i(z_1,..,z_L;\zeta_l (z_0), \zeta_r(z_{L+1}))=\prod_{0\leq j\neq i\leq L+1}
\frac{(z_i+z_j)(z_i z_j+1)}{z_i z_j},
\end{align}
satisfies  (\ref{mubqKZ3}) with $i=1$, provided we choose appropriately the 
constant $\psi_{\pi_0}^*$. Although we computed (\ref{factor}) for 
$i=1$, it is also true for $i>1$. One can easily see this by combining this recurrence relation with the $q$KZ equation. 

\subsection{Computation of the \texorpdfstring{$\psi_{\pi_i}$}{Lg} components}
So far we found only the fully nested elements $\psi_{\pi_0}$ and $\psi_{\tilde{\pi}_0}$. 
The only $q$KZ equation that relates the fully nested element with others is (\ref{bqKZ2}), 
however, it has two unknowns: $\psi_{\{1,-1..,-1\}}$ and $\psi_{\pi_1}=\psi_{\{0,-1,..,-1\}}$. 
As we already saw, after the action of $\mu_1$ one of the two unknowns in (\ref{bqKZ2}) 
disappear and we arrive at (\ref{mubqKZ2}), hence we can express the element $\psi_{\pi_1}$ 
in terms of $\psi_{\pi_0}$ and (\ref{bqKZ2}) gives us the second unknown 
$\psi_{\{1,-1..,-1\}}$. 

\begin{figure}[htb]
\centering
\includegraphics[width=0.55\textwidth]{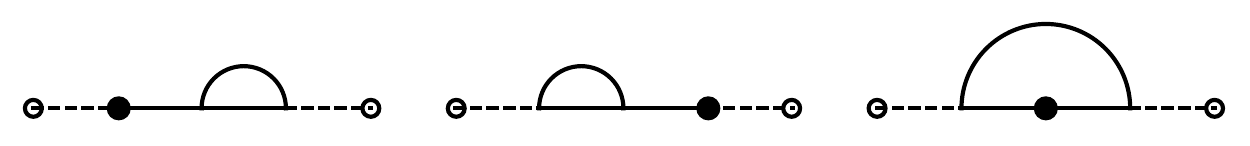}
\caption{Equivalence class of the element $\pi=\{0,1,-1\}$.}
\label{figeq1m1}
\end{figure}

In general, (\ref{mubqKZ2}) holds if we replace $\pi_0$ 
by $\pi=\{-1,-1,\alpha\}$ for any $\alpha$. Using (\ref{mubqKZ2}) we can map the 
component $\psi_{\{-1,-1,\alpha\}}$ and $\psi_{\{-1,\alpha\}}$ of the ground state 
$\Psi_{L+1}$ and $\Psi_{L}$ (respectively) to the components $\psi_{\{0,\alpha\}}$ of the 
ground state $\Psi_{L}$:
\begin{align}\label{0alpha}
\psi_{\{0,\alpha\}}(z_1,z_2,..,z_L))=&k_{l,4}^{-1}(z_1\omega,\zeta_l)\bigg{( }
U_l(z_1\omega,\zeta_l)\frac{\psi_{\{-1,-1,\alpha\}}(1/(z_1 \omega),z_1/\omega,z_2,..,z_L)}
{F_1(z_1,..,z_L)}+\nonumber \\
&k_{l,1}(z_1\omega,\zeta_l)\psi_{\{-1,\alpha\}}(z_1,z_2,..,z_L)\bigg{)}.
\end{align}
This equation is the first important ingredient in the computation of $\psi_{\pi_i}$. 
The second ingredient is (\ref{qKZhop1}) which allows one to 
compute the component $\psi_{\{\alpha,0,\pm 1,\beta\}}$ if the component 
$\psi_{\{\alpha,\pm 1,0,\beta\}}$ is known and vice versa.  The latter means all components 
with a fixed relative connectivity of the lines (relative positions of $+1$'s and $-1$'s in
$\pi$) form an equivalence class w.r.t. the action of the operators (see Fig. \ref{figeq1m1})
\begin{align}\label{hop}
&\eta_i=\frac{W(z_i,z_{i+1})}{r_5(z_i,z_{i+1})}(\tau_i-1)+1,\\
& \text{where}~~~\tau_i f_(..,z_i,z_{i+1},..)=f(..,z_{i+1},z_i,..), \nonumber
\end{align}
thus, we get the identities:
\begin{align}
\psi_{\{\alpha,\pm 1_i,0_{i+1},\beta\}}=\eta_i \psi_{\{\alpha,0_i,\pm 1_{i+1},\beta\}},\label{hop21}\\
\psi_{\{\alpha,0_{i},\pm 1_{i+1},\beta\}}=\eta_i \psi_{\{\alpha,\pm 1_{i},0_{i+1},\beta\}}.\label{hop22} 
\end{align}
Plugging $\eta_1 \psi_{\pi_1}(z_1,..,z_{L})$ and $\eta_2\eta_1 \psi_{\pi_1}(z_1,..,z_{L+1})$ 
into (\ref{0alpha}) gives $\psi_{\pi_2}$. It is clear now that by substituting:
\begin{align}
&\psi_{\{-1,\alpha\}}(z_1,..,z_L)=\prod_{k=0}^{i-1}\eta_{i-k}\psi_{\pi_i}(z_1,..,z_L)
\label{pii1},\\
&\psi_{\{-1,-1,\alpha\}}(z_1,..,z_{L+1})=
\prod_{k=0}^{i-1}\eta_{i-k+1}\eta_{i-k}\psi_{\pi_i}(z_1,..,z_{L+1}), \label{pii2}
\end{align}
into (\ref{0alpha}) we can inductively obtain all $\psi_{\pi_i}$. Note, that the 
 (\ref{bqKZ2}) for $\pi=\{-1,-1,\alpha\}$ where $\alpha$ contains $0$'s and/or $-1$'s 
allows one to obtain all elements $\psi_{\pi}$ with $\pi=\{1,-1,\alpha\}$. 
Using $\eta_i$'s we also get all elements in the equivalence of $\psi_{\{1,-1,\alpha\}}$. 
Now we can turn to the computation of the other elements of the ground state $\Psi_L$. 
\begin{figure}[htb]
\centering
\includegraphics[width=0.6\textwidth]{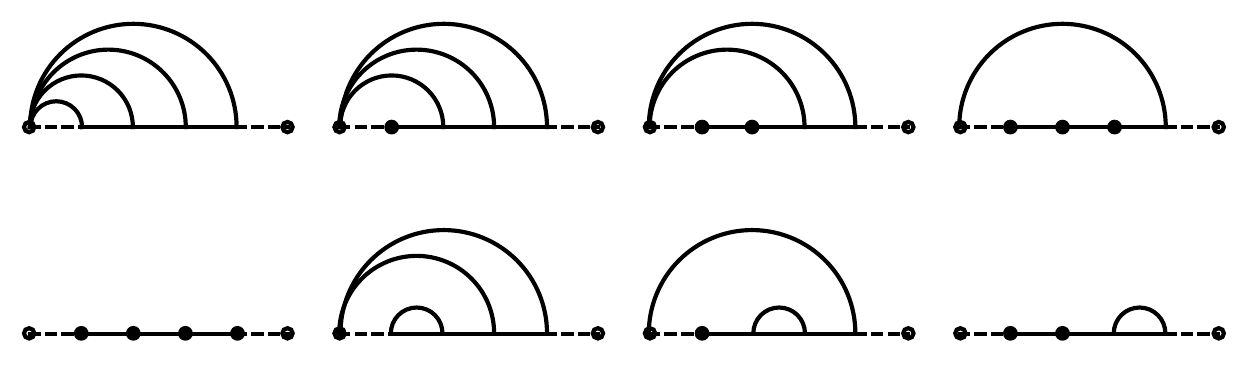}
\caption{Elements corresponding to nonequivalent link patterns of  $\Psi_{4}$ can be computed using only s 
 (\ref{mubqKZ2}), (\ref{bqKZ2}), (\ref{pii1}) and (\ref{pii2}).}
\label{figlp1}
\end{figure}

\subsection{Dyck paths and the ground state components}
\begin{figure}[htb]
\centering
\includegraphics[width=0.65\textwidth]{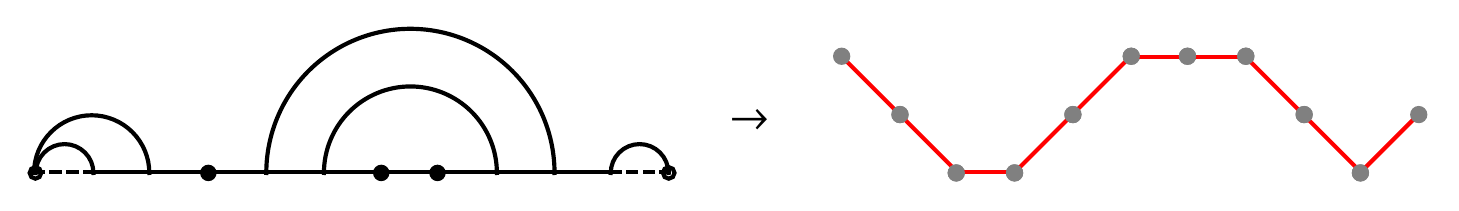}
\caption{The link pattern $\{-1,-1,0,1,1,0,0,-1,-1,1\}$ and the corresponding Dyck path.
The ending and the beginning of each step in the Dyck path is marked by a small disk on 
the picture.}
\label{figDP}
\end{figure}

\begin{figure}[htb]
\centering
\includegraphics[width=0.3\textwidth]{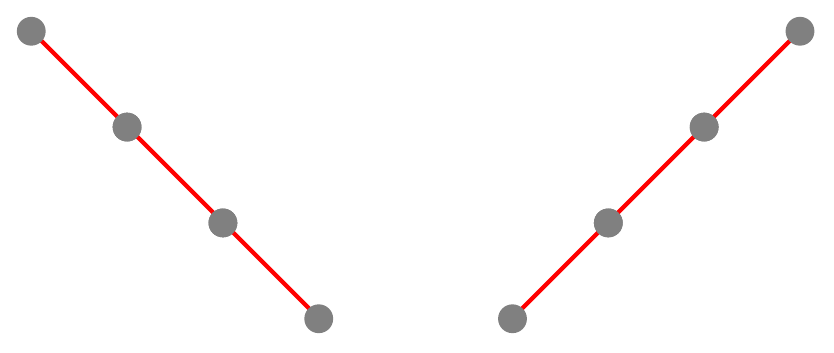}
\caption{The Dyck paths of $\pi=\{-1,-1,-1\}$ and $\tilde{\pi}=\{1,1,1\}$.}
\label{figDP1}
\end{figure}

\begin{figure}[htb]
\centering
\includegraphics[width=0.6\textwidth]{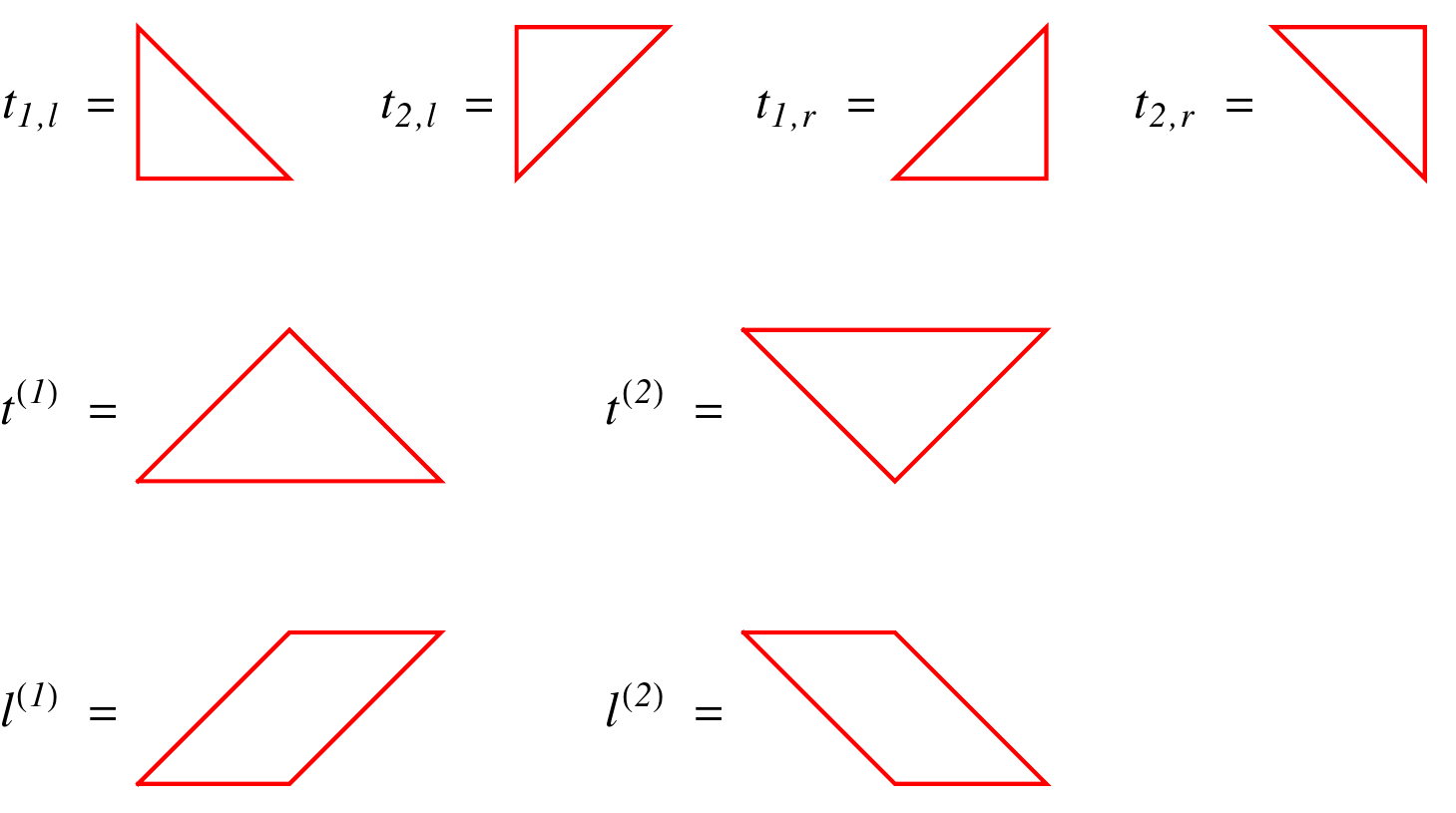}
\caption{The left boundary $t_{1,l}$, $t_{2,l}$, right boundary $t_{1,r}$, $t_{2,r}$,
the bulk $t^{(1)}$, $t^{(2)}$ triangles and the bulk rhombi $l^{(1)}$, $l^{(2)}$.}
\label{figtriang}
\end{figure}

\begin{figure}[htb]
\centering
\includegraphics[width=0.5\textwidth]{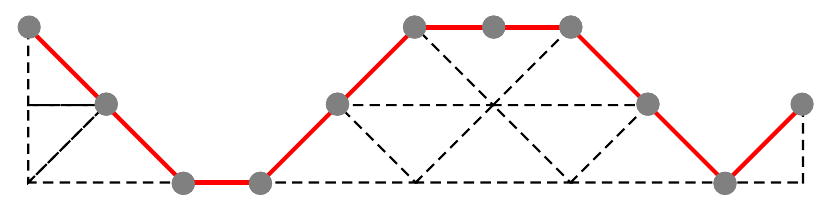}
\caption{Onepossible decomposition of the Dyck path $\{-1, -1, 0, 1, 1, 0, 0, -1, -1, 1\}$.}
\label{figDycktriang}
\end{figure}

As in \cite{PDF1}, a good way to see how the components of $\Psi_L$ can be computed is to turn to 
the Dyck path formulation.  The set of link patterns of length $L$ is the set 
$\mathbb{Z}_3^{\otimes L}$.  To any $\pi$ we associate a unique Dyck path in 
the following way. Reading $\pi$ from left to right, every $1$ corresponds to 
the north east (NE) step, every $-1$ corresponds to the south east (SE) step and 
every $0$ corresponds to the horizontal east (E) step. For example, the link pattern 
$\{-1,-1,0,1,1,0,0,-1,-1,1\}$ is represented by the sequence of moves \{SE,SE,E,NE,NE,E,E,SE,SE,NE\} which is a Dyck path given in Fig. \ref{figDP}.  
A few distinguished Dyck paths are those which correspond to the 
fully nested link patterns (see Fig. \ref{figDP1}). 
The area under the Dyck paths can be split (in a non unique way) into a few boundary triangles $t_{l,1}$, $t_{l,2}$, 
$t_{r,1}$ and $t_{r,2}$, bulk triangles $t^{(1)}$ and $t^{(2)}$ and lozenges $l^{(1)}$ and $l^{(2)}$ 
depicted in Fig.   \ref{figtriang}. Note, that we need to supplement the bulk triangles
and lozenges with an index $j$, then by $t^{(i)}_j$ and $l^{(i)}_j$ we will denote the 
operator $t^{(i)}$, respectively $l^{(i)}$, acting at the position $j$.  
Each Dyck path can be build by concatenation of these primitives starting from a 
reference Dyck path. For example, if the reference Dyck path is $\pi_e=\{0,0,..,0\}$
the Dyck path $\pi=\{-1,-1,0,1,1,0,0,-1,-1,1\}$, depicted on the 
Fig. (\ref{figDycktriang}), is constructed, e.g. as follows
\begin{align}
\pi=t_{6}^{(2)}t_{7}^{(1)}t_{5}^{(1)}t_{1,l}t_{7}^{(2)}t_{5}^{(2)}t_{2,l}
t_{1}^{(1)}t_{4}^{(1)}t_{6}^{(1)}t_{8}^{(1)}t_{1,r}\pi_e.
\end{align}
Similarly, as one link pattern can be converted into another upon the action of a product of 
$\rho_i$ operators, one Dyck path can be converted into another one by attaching or 
removing rhombi $l^{(i)}$ and triangles $t^{(i)}$ and $t_{i,l/r}$. 

\begin{figure}[htb]
\centering
\includegraphics[width=0.15\textwidth]{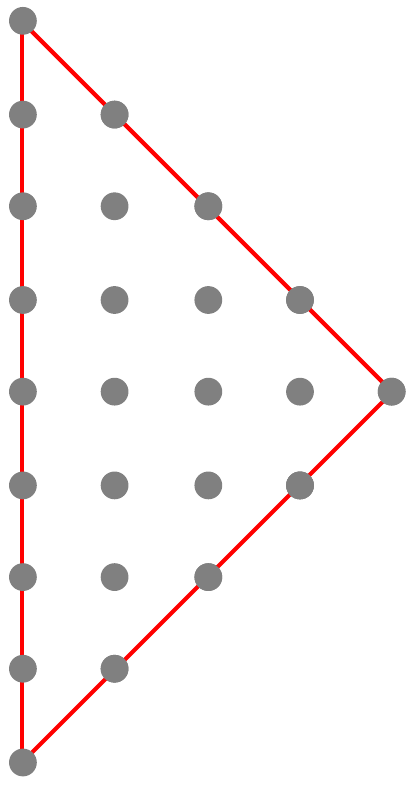}
\caption{The domain $D_L$.  The gray disks represent the nodes which the possible Dyck paths may 
visit.}
\label{figregion}
\end{figure}

Now we introduce the notion of partial ordering on the Dyck Paths. A path $\pi_1$ is contained 
in another path $\pi_2$: $\pi_1\prec\pi_2$ if $\pi_1$ can be completed to $\pi_2$ by adding 
to it triangles and lozenges.  To make this 
notion less ambiguous we need to indicate, for example, what is the largest Dyck path 
w.r.t.$~$all other Dyck paths. Let us choose the element $\pi_0$ to be the largest, then the 
smallest element will be $\tilde{\pi}_0$.  This means that the addition of the triangles or lozenges 
to the element $\pi_0$ or the removal of the triangles or lozenges from the element $\tilde{\pi}_0$ is forbidden. Hence all Dyck paths are contained in the region surrounded by the triangle 
$D_L=\{\{0,L\},\{L,0\},\{0,-L\}\}$.  This region $D_L$ is depicted in Fig.   \ref{figregion}.    

In order to compute the remaining elements of the ground state we will need the 
 (\ref{qKZem}), which we rewrite in the following form for $\pi=\{\alpha,0,0,\beta\}$:
\begin{align}
&\delta_i \psi_{\{\alpha,0,0,\beta\}}(..,z_{i},z_{i+1},..)-
\psi_{\{\alpha,1,-1,\beta\}}(..,z_{i},z_{i+1},..)=
\sum_{\substack{
           \pi^{\prime}: \rho_i^{(8)} \pi^{\prime}=\pi\\
            \pi^{\prime}\neq \pi}}
\psi_{\pi^{\prime}}(..,z_i,z_{i+1},..)\label{d2},
\end{align}
where we defined the operator $\delta_i$:
\begin{align}\label{delta}
\delta_i=\frac{W(z_i,z_{i+1})\tau_i-r_7(z_i,z_{i+1})}{r_1(z_i,z_{i+1})}.
\end{align}
The boundary $q$KZ with $\pi=\{0,\alpha\}$  (\ref{bqKZ3}) is also useful to rewrite as:
\begin{align}
&\gamma_l\psi_{\{0,\alpha\}}(z_1,..)-\psi_{\{1,\alpha\}}(z_{1},..)=
\psi_{\{-1,\alpha\}}(z_{1},..),\label{b1}
\end{align}
where we defined the operator $\gamma_l$:
\begin{align}
\gamma_l=\frac{U_l(z_1,\zeta_l)\sigma_l-k_{l,3}(z_1,\zeta_l)}{k_{l,2}(z_1,\zeta_l)},~~~~ 
\text{and}~~\sigma_l f(z_1,z_2,..)=f(1/z_1,z_2,..).
\end{align}
Note, in both equations  (\ref{d2}) and  (\ref{b1}) in the right hand sides we can identify
the smallest elements in those equations in the sense of the 
ordering of Dyck paths.  The idea below is to compute one by one elements $\psi_{\pi}$ in the 
direction of decreasing $\pi$, therefore  (\ref{d2}) and  (\ref{b1}) will always 
have one unknown which will be the element corresponding to the minimal $\pi$ within a given equation.

\begin{figure}[htb]
\centering
\includegraphics[width=1\textwidth]{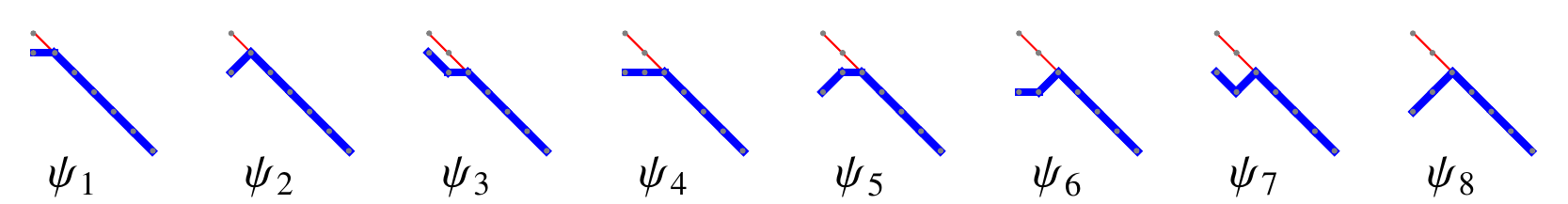}
\caption{The Dyck paths (thick blue line) passing through the point $\{1,L-1\}$ ($\psi_1$ and 
$\psi_2$) and 
the Dyck paths passing through the point $\{2,L-2\}$ ($\psi_j,~~j=3,..,8$) of the domain $D_L$. For the reference we also included the north to east edge of the $D_L$ triangle.}
\label{figdpaths1}
\end{figure}

We proceed as follows. Let us consider the elements corresponding to the 
Dyck paths passing through the points $\{j,L-j\}$ for $0\leq j\leq L $, i.e. the points on the right 
boundary of the domain $D_L$.  Take $j=0$, there is only one path which passes through 
this point and it corresponds to $\psi_{\pi_0}$. Next we take $j=1$, there 
are two paths corresponding to $\psi_1$ and $\psi_2$ in Fig.   \ref{figdpaths1}. 
Next we take $j=2$, these paths are labeled by $\psi_i$ with $i=3,..,8$ on the 
Fig.  (\ref{figdpaths1}). We already know the following components: $\psi_0$, $\psi_1$, $\psi_2$, $\psi_3$ 
and $\psi_4$. Here, $\psi_0$ is the fully nested component (\ref{fnfactor}), while the others follow from (\ref{0alpha}), (\ref{hop21}), (\ref{hop22}) and  (\ref{b1}). 
\begin{figure}[htb]
\centering
\includegraphics[width=1\textwidth]{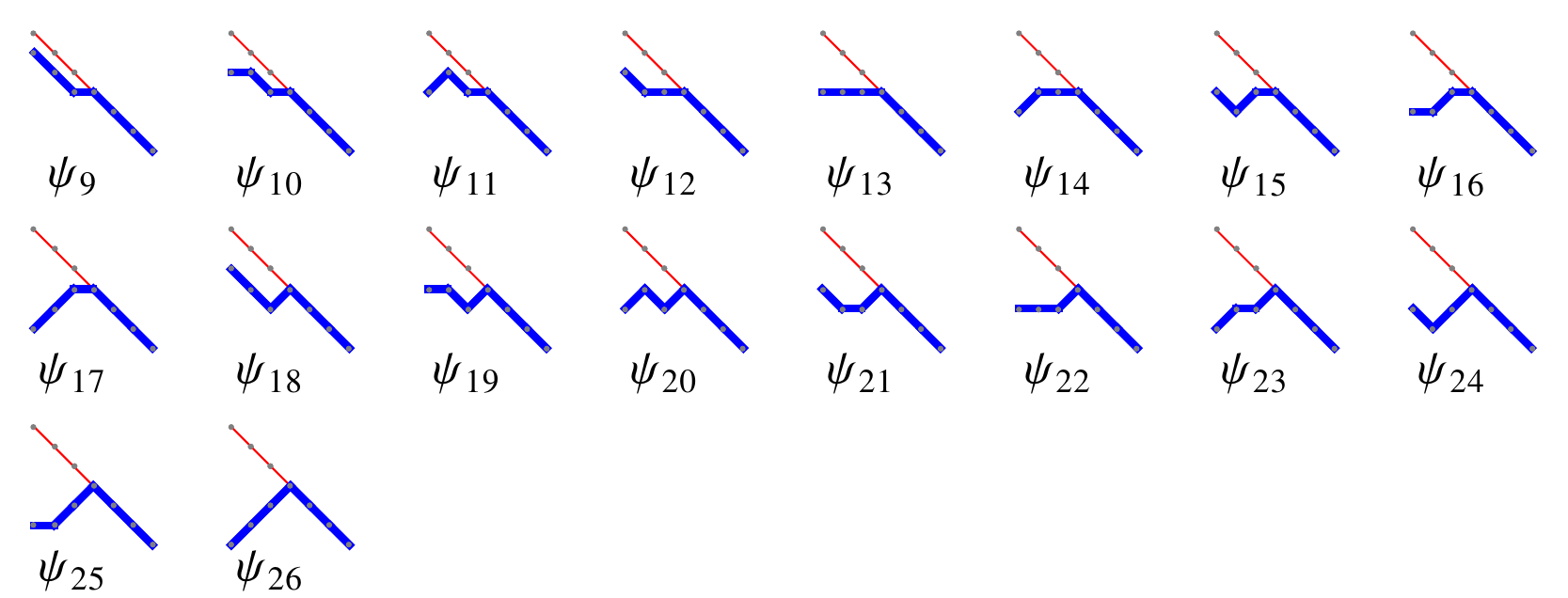}
\caption{Dyck paths passing the $\{3,L-3\}$ point of the domain $D_L$.}
\label{figdpaths2}
\end{figure}
Every unknown $\psi_i$ on this figure we can obtain from a few $\psi_j$'s with $j<i$ and 
$\psi_0=\psi_{\pi_0}$. Indeed, we have:  
\begin{align}\label{deltapsi}
&\psi_5=\gamma_l \psi_4 -\psi_3,~~~
\psi_6=\eta_1 \psi_5,~~\psi_7=\delta_1 \psi_4-\psi_2-\psi_0,~~~
\psi_8=\gamma_l \psi_6-\psi_7.
\end{align}
Let us take now $j=3$, the Dyck paths passing through this point are depicted on the 
Fig. \ref{figdpaths2}. 
Among these we already know the components which are in the equivalence class w.r.t. the action of $\eta$ of $\psi_{\pi_i}$ and $\psi_i$ from Fig.  \ref{figdpaths1}.  These components are: $\psi_9$, 
$\psi_{10}$, $\psi_{11}$, $\psi_{12}$, $\psi_{13}$, the rest are:
\begin{align}\label{deltapsi2}
&\psi_{14}=\gamma_l \psi_{13} -\psi_{12},~~~
\psi_{15}=\delta_1 \psi_{13}-\psi_{11}-\psi_9,~~~
\psi_{16}=\eta_1 \psi_{14},~~~
\psi_{17}=\gamma_l \psi_{16}-\psi_{15},~~~\nonumber\\
&\psi_{18}=\delta_2 \psi_{12}-\psi_{7}-\psi_{0},~~~
\psi_{19}=\eta_1 \eta_2 \psi_{15},~~~
\psi_{20}=\gamma_l \psi_{19} - \psi_{18},~~~
\psi_{21}=\eta_1 \psi_{15},~~~\nonumber\\
&\psi_{22}=\eta_2 \eta_1 \psi_{14},~~~
\psi_{23}=\gamma_l \psi_{22}-\psi_{21},~~~
\psi_{24}=\delta_1 \psi_{22}-\psi_{18}-\psi_{20}-\psi_{8},~~~\nonumber\\
&\psi_{25}=\eta_1 \eta_2 \psi_{17},~~\psi_{26}=\gamma_l \psi_{25}-\psi_{24}.
\end{align}
These computations are valid for any $L$ larger or equal to $3$. In the latter case it gives 
all 27 components. In this calculation we have selected the reference state 
to be $\psi_{\pi_0}$.  Then we found that 
any component $\psi_{\pi}$ can be expressed through a number of components associated 
to larger Dyck paths than the path $\pi$. All computations can be done in a similar manner 
using $\psi_{\tilde{\pi}_0}$ as the reference.  The explicit results for $L=1,2$ are 
presented in Appendix \ref{appA}.

\section{Discussions}\label{sec4}
Usually, integrable models at finite size are studied via the Bethe ansatz, in which case 
the eigenstates of the transfer matrix depend on the Bethe roots. One then has to solve 
the set of nonlinear Bes in order to obtain the eigenstates and 
the eigenvalues. In our case, 
following the prescription developed for the TL model at $n=1$ \cite{PDFPZJ}, we were able to find 
the ground state eigenvector explicitly for finite systems avoiding the complicated nonlinear 
equations. Indeed, if we take 
our ground state $\Psi_L$ in the spin basis (which amounts to a linear transformation on 
$\Psi_L$, see e.g. \cite{Nienhuis1}) this will correspond to a certain state of the 
19 vertex Izergin--Korepin (IK) model  at $q=e^{i\pi/3}$.  The algebraic Bethe ansatz for 
this model with periodic boundary conditions was found by Tarasov \cite{Tarasov}. In the case of periodic boundary conditions and small system sizes we verified that the Bethe ansatz eigenstate agrees with the calculation based on the $q$KZ equation of the type presented in this paper.

The approach that we are using allows us to find (strictly speaking to conjecture) closed 
expressions for a few components of the ground state vector.  These 
components are the fully nested element $\psi_{\pi_0}$ and a few others which are ``close" 
enough to $\psi_{\pi_0}$ and expressed through $\psi_{\pi_0}$ itself. For example  
$\psi_{\pi_1}$ follows from (\ref{0alpha}). 
As we go further away from these components the complexity of the computations increases rapidly, 
that is why we are limited to only a few components. Another component which we 
can write in a closed form for arbitrary $L$ is $\psi_e$. It turns out that this component 
is proportional to $Z_L$-the sum of all components of $\Psi_L$. $Z_L$ is the 
normalization of the ground state $\Psi_L$, thus it is an important quantity for the 
computation of correlation functions. One such correlation function is the boundary to 
boundary current. It was computed for the TL model in \cite{dGNP} and in the dilute TL case it is 
the subject of the paper \cite{FN}. Due to the relation of the dilute TL model at $n=1$ 
to the critical site percolation, $Z_L^2$ plays the role of the partition function in 
the critical percolation on an infinite strip. We compute $Z_L$ in \cite{GN}.

We would like to stress here that we were dealing with the generic open boundary 
conditions. 
In this case the boundaries carry the parameters $\zeta_l$ and $\zeta_r$. 
From our state  $\Psi_L$ we can recover the 
ground state $\Psi_L^c$ of the loop model with closed (also called reflecting) boundary 
conditions simply by sending the parameters $\zeta$ to $0$ (where  $\Psi_L$ is assumed to be properly normalized). In this case one could study an interesting correlation function: the 
current going from $-\infty$ to $+\infty$ along the strip. 
This current was studied in the paper \cite{IP} for the TL $n=1$ 
loop model. One could also send $\zeta$ to infinity (with a proper normalization), then 
the corresponding $K$-matrix is a combination of $\kappa_3$, $\kappa_4$ and $\kappa_5$.
This can be an interesting case to study, since as it is stressed in \cite{dGLR} there are a 
few solutions $K^{(1)}$ and $K^{(2)}$ to the reflection equation.  The solution $K^{(1)}$ is 
the one we used here.  The solution $K^{(2)}$ is a combination of 
only three operators $\kappa_3$, $\kappa_4$ and $\kappa_5$.  The two solutions 
are independent for generic $n$, but for $n=1$ $K^{(2)}$ is the $\zeta\rightarrow \infty$ 
limit of $K^{(1)}$. Hence to compute the ground state 
with $K^{(2)}$-matrix one needs to send $\zeta$ to infinity in our solution.

The next point we would like to stress is the computation of the ground state at $n=0$. 
It turns out that at $n=0$ the dilute TL loop model also possesses a polynomial 
ground state and the $q$KZ together with the recurrence relations remain intact. 
The algorithm of the computation of $\Psi_L$ presented here remains applicable. 

\section*{Acknowledgements}
A. G. was supported by the AMS Amsterdam Scholarship, ERC grant 278124 "Loop models, Integrability 
and Combinatorics" and the Australian Research Council Centre of Excellence for Mathematical and Statistical Frontiers. A. G. thanks the University 
of Amsterdam for hospitality and support and P. Zinn-Justin and G. Feh\'{e}r for useful discussions.   

\appendix

\section{Appendix}\label{appA}
In this appendix we present explicit expressions for the polynomial ground state 
$\Psi_L$ for $L=1,~2$.  The $\Psi_1$ components are:
\begin{align}
&\psi_{-1}=\frac{\omega  \left(x_r+\omega  z_1\right) 
\left(\omega  z_1 x_r+1\right)}{z_1 x_r}, \label{psim1} \\
&\psi_{1}=\frac{\omega  \left(\omega  x_l+z_1\right) 
\left(z_1 x_l+\omega \right)}{z_1 x_l}, \label{psi1} \\
&\psi_{0}=\frac{\omega ^2 \left(z_1 x_l^2 x_r+z_1^2 x_l x_r+z_1 x_l x_r^2+
x_l x_r+z_1 x_l+z_1   x_r\right)}{z_1 x_l x_r}. \label{psi0}
\end{align}
The last component can be written in terms of the elementary symmetric polynomials 
\begin{align}
E_{m}(z_{1},..,z_{L})&=\sum_{1\leq i_{1}<,..,<i_{m}}z_{i_{1}}..z_{i_{m}}, \nonumber \\
E_{m}(z_{1},..,z_{L})&=0~~~ \text{for}~~~ m<0,~~~ \text{and}~~~ m>L,\nonumber 
\end{align}
as follows:
\begin{align}
\psi_{0}(z_1;x_l,x_r)=\frac{\omega ^2(E_2+
E_1 E_3)}{E_3},
\end{align}
in which $E_i=E_i(x_l,z_1,x_r)$.
The choice of the constant prefactor of the nested elements in  (\ref{psim1}) and  (\ref{psi1}) 
fixes the constants $\psi_{\pi_0}^*$ and $\psi_{\tilde{\pi}_0}^*$ from  (\ref{fnfactor}) 
and  (\ref{fntfactor}). Hence we will not write the nested elements for $L=2$. 
The remaining components of $\Psi_2$ up to the actions of $\eta_i$ are:
\begin{align}
&\psi_{1,-1}=\frac{\left(\omega  x_l+z_1\right) \left(z_1 x_l+\omega \right) \left(x_r+\omega 
   z_2\right) \left(\omega  z_2 x_r+1\right)}{z_1^2 z_2^2 x_l^2 x_r^2} (E_3 +E_1 E_4),
\nonumber \\
&\psi_{0,-1}=
\frac{\omega  \left(x_r+\omega  z_2\right) \left(\omega  z_2 x_r+1\right)}
{z_1^2 z_2^2 x_l x_r^2} \bigg{(}z_1(1+z_2^2) (E_2(x_l,z_1,x_r)+ 
E_1(x_l,z_1,x_r) E_3(x_l,z_1,x_r))+ \nonumber\\
&z_2 \big{(}\left(z_1^2+1\right) x_l \left(x_r+z_1\right) \left(z_1 x_r+1\right)+ 
z_1 (x_1^2+1)\left((\omega+1) x_r \left(\omega^2+z_1^2\right)+z_1( x_r^2+1)\right)\big{)}\bigg{)}, 
\nonumber\\
&\psi_{1,0}=
\frac{\omega \left(\omega x_l+z_1\right) \left(z_1 x_l+\omega\right)}{z_1^2 z_2^2 x_l^2 x_r}
\bigg{(}
z_2(1+z_1^2) (E_2(x_l,z_2,x_r)+ E_1(x_l,z_2,x_r) E_3(x_l,z_2,x_r))+ \nonumber\\
&z_1 \big{(}z_2 \left(x_l^2+1\right) \left(x_r+z_2\right) \left(z_2 x_r+1\right)+
x_l \left((\omega+1) z_2 \left(x_r^2+1\right) \left(\omega^2
   z_2^2+1\right)+\left(z_2^2+1\right){}^2 x_r\right)     \big{)}
\bigg{)}\nonumber,
\end{align}
\begin{align}
&\psi_{-1,1}=\frac{\omega (\omega+1) \left(z_1 z_2+1\right) \left(\omega z_1+z_2\right)}
{z_1^2 z_2^2 x_l x_r}\bigg{(} (\omega +1) x_l x_r (1+z_1^2 z_2^2)+\nonumber\\ 
&(z_1+z_2)(\omega  x_l x_r^2+x_l^2 x_r+\omega  x_l+x_r)+
(\omega +1) z_1 z_2 \left(x_l^2 x_r^2+x_l x_r+x_l^2+x_r^2+1\right)+\nonumber\\
&z_1 z_2 \left(z_1+z_2\right)\left(\omega  x_l^2 x_r+x_l x_r^2+x_l+\omega 
   x_r\right) \bigg{)},\nonumber\\
&\psi_{0,0}=\frac{\omega^2}{E_4^2}(E_3 +E_1 E_4)(E_2-E_4+E_1 E_3+E_2 E_4).
\end{align}
In these equations when the arguments of $E_i$ are not specified we imply 
$E_i=E_i(x_l,z_1,z_2,x_r)$.

\section{Appendix}\label{appB}
\begin{figure}[htb]
\centering
\includegraphics[width=1\textwidth]{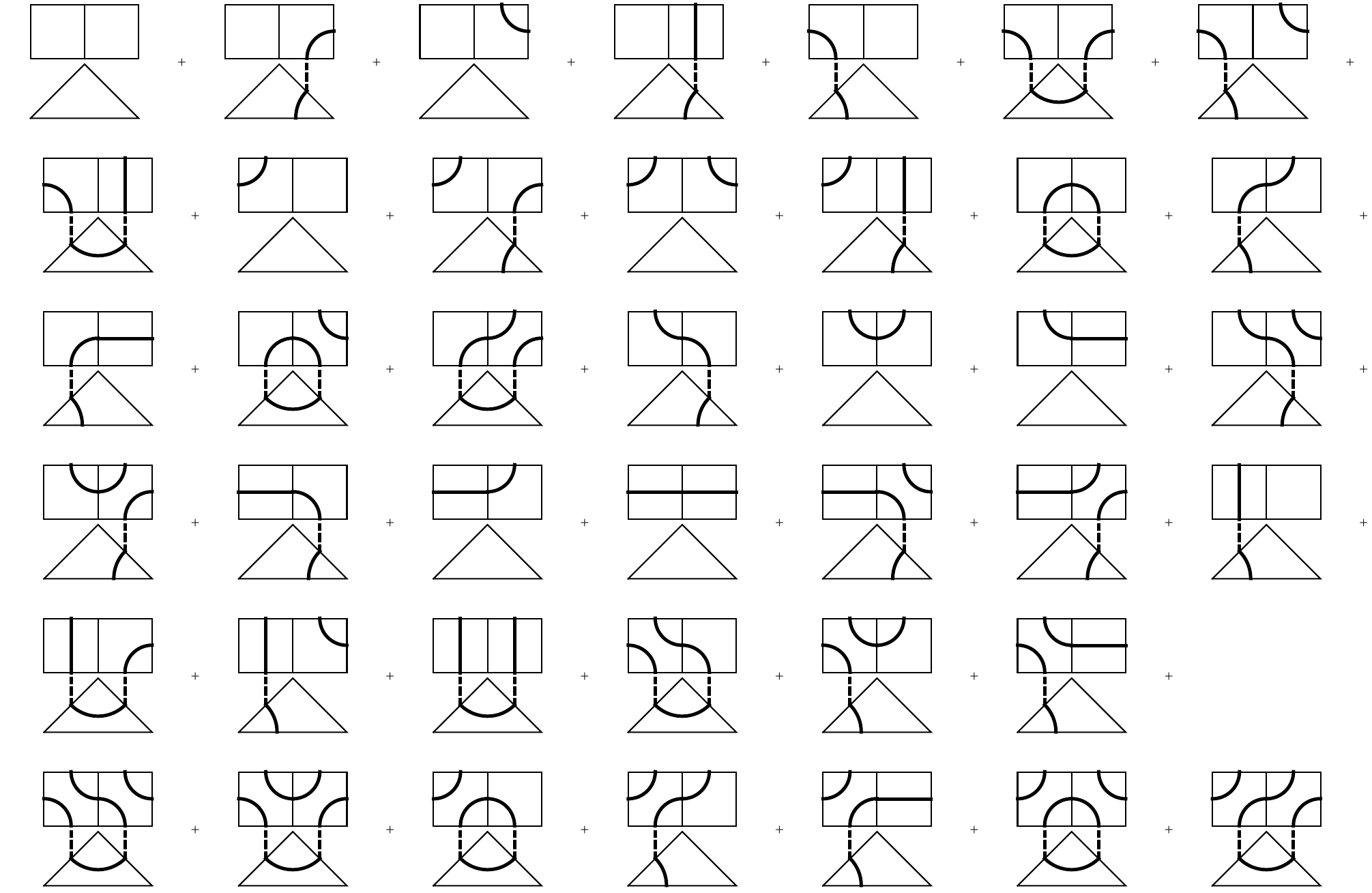}
\caption{The left hand side of  (\ref{figYenB}).  The weights of each term here 
are the products of the weights of the constituting operators.}
\label{LHSY}
\end{figure}

\begin{figure}[htb]
\centering
\includegraphics[width=0.8\textwidth]{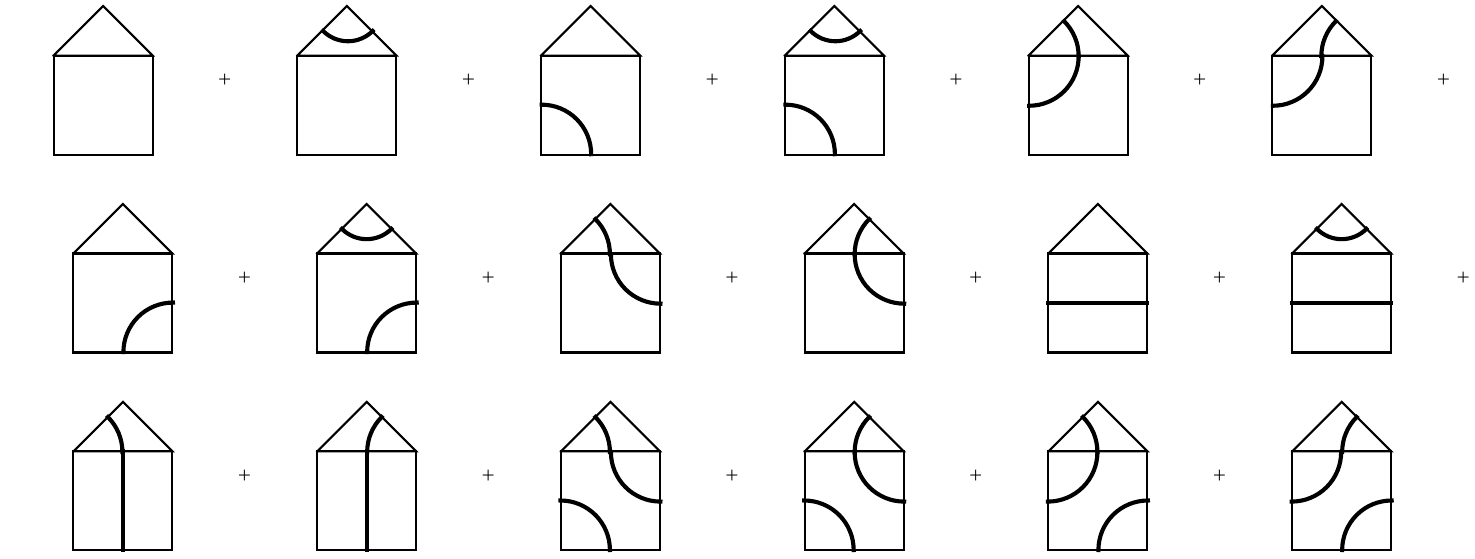}
\caption{The right hand side of   (\ref{figYenB}).  The weights of each term 
here are the products of the weights of the constituting operators.}
\label{RHSY}
\end{figure}

In this appendix we discuss in more detail   (\ref{MRR}) (see also Fig. \ref{figYenB}). When the argument of the 
matrix $R(z)$ is equal to $\omega^2$ the weight $\rho_9$ in  (\ref{Rweights1}) vanishes, and the $R$-matrix can be split 
into a product of two operators shown in Fig.  \ref{figMS}. Now we take $M$-operator and substitute it into  
Fig. \ref{figYenB}. 
The left hand side of this equation is shown in Fig.  \ref{LHSY}, while the right hand side is shown in Fig.  \ref{RHSY}.
On both sides we have five external edges.  The sum of the weights of the terms with the same connectivity on the left hand side must 
be equal to the sum of the weights of the terms with the same connectivity on the right hand side times the factor $(t^2-z_i^2)$. 
For example, 
take on the left hand side of the equation the terms with all five external edges empty: the one with no lines 
in the interior of the diagram and the one with a loop in the interior.  This will give the weight $(t^2-z_i^2)^2$ which is equal to the weight $\rho_7(t,z_i)$ times the factor $(t^2-z_i^2)$.  The latter is in agreement with the computation on the right hand side. 
In particular, all terms on the left hand side which have two top 
horizontal edges occupied and not connected cancel each other since such a connectivity is absent on the right 
hand side. It is a straightforward but a lengthy computation to check that  (\ref{MRR}) holds. 

The factorization property of the $R$-matrix holds for generic values of $q$. In this case the 
empty plaquette of the $M$-operator will 
acquire the weight of $n$ and the same for the $S$-operator.  The derivation of this 
fact will appear elsewhere \cite{Ga}.
In the context of quantum 
groups this factorization of the $R$-matrix is related to the quasi-triangularity 
property which is one of the defining properties of the quasi-triangular Hopf algebras \cite{Mont}.

\small{\bibliographystyle{plain}}
\bibliography{bibgs}

\end{document}